\documentclass[
    ,draft            
  ]
  {aipproc}

\layoutstyle{6x9}

\def\boldmath{}
\def\mevm{~MeV/$c^2$\/}

\def\meve{~MeV}
\def\gevm{~GeV/$c^2$\/}
\def\gevp{~GeV/$c$\/}
\def\geve{~GeV}
\def\simge{\mathrel{%
   \rlap{\raise 0.511ex \hbox{$>$}}{\lower 0.511ex \hbox{$\sim$}}}}
\def\simle{\mathrel{
   \rlap{\raise 0.511ex \hbox{$<$}}{\lower 0.511ex \hbox{$\sim$}}}}
\def\belle{Belle}
\def\cherenkov{\u{C}herenkov}

\def\qqbar{$q\bar{q}$}
\def\Rqq{$R^{}_{q\bar{q}}$}
\def\bhh{$B\!\rightarrow\!hh$}
\def\bpipi{$B^0\!\rightarrow\!\pi^+\pi^-$}
\def\bomegak{$B\!\rightarrow\!\omega K$}
\def\bomegapi{$B\!\rightarrow\!\omega\pi$}
\def\acp{$A^{}_{CP}$}

\def\phikk{$\phi\rightarrow K^+K^-$}
\def\cp{$CP$}
\def\bphik{$B\rightarrow\phi K$}
\def\bphikst{$B\rightarrow\phi K^*$}
\def\azero{$A^{}_0$}
\def\apar{$A^{}_\parallel$}
\def\aperp{$A^{}_\perp$}
\def\thetatr{$\theta^{}_{tr}$}
\def\phitr{$\phi^{}_{tr}$}
\def\thetakst{$\theta^{}_{K^*}$}
\def\hzerzer{$H^{}_{00}$}
\def\honeone{$H^{}_{11}$}
\def\brhorho{$B^+\rightarrow\rho^+\rho^0$}
\def\rhopipi{$\rho\rightarrow\pi\pi$}
\def\thetahel{$\theta^{}_{hel}$}
\def\bdcpk{$B^\pm\rightarrow D^{}_{CP}K^\pm$}
\def\dcp{$D^{}_{CP}$}

\begin{document}

\begin{flushright}
UCHEP-03-01
\end{flushright}
\vskip-0.30in

\title{Rare hadronic $B$ decays: probing deeper into the Standard Model}

\author{A.\,J.\,Schwartz}{
  address={Physics Department, University of Cincinnati, Cincinnati, Ohio 45221}
}

\begin{abstract}
We present recent results from the \belle\ experiment 
on rare hadronic $B$ meson decays. The results are 
based on a 78~fb$^{-1}$ data sample and consist of 
branching fractions, \cp\ asymmetries, and polarization
amplitudes. The decays studied include two-body 
pseudoscalar-pseudoscalar final states
($B\rightarrow\pi\pi,\ K\pi,\ KK$, and $D^0 K^\pm$); 
pseudoscalar-vector final states 
($B\rightarrow\omega\pi,\,\omega K$, and $\phi K$); and
vector-vector final states 
($B\rightarrow\phi K^*$ and $\rho^+\rho^0$).
\end{abstract}

\maketitle

\section{Introduction}

Rare hadronic $B$ decays are useful for probing physics beyond the
Standard Model. Their amplitudes usually contain internal
loops, which are sensitive to mass scales that cannot
be accessed directly. Here we present recent measurements 
of such decays by the \belle\ experiment at KEK. 
This experiment runs at the KEKB asymmetric $e^+e^-$
collider, which has a center-of-mass (CM) energy near the $\Upsilon(4S)$
resonance. The results are from 78~fb$^{-1}$ of data, which
corresponds to $85\times 10^6$ $B\overline{B}$ pairs produced.

The \belle\ detector consists of a three-layer 
silicon vertex detector, a 50-layer central drift chamber
(CDC) for charged-particle tracking, an array of silica 
aerogel threshold \cherenkov\ counters (ACC),
time-of-flight scintillation counters (TOF), a CsI(Tl) 
electromagnetic calorimeter (ECL), and a superconducting 
solenoid providing a 1.5 T magnetic field. An 
iron flux-return located outside the coil is instrumented
with resistive plate chambers to identify muons and $K^0_L$'s.
For details of the detector, see Ref.~\cite{detector}.

Charged tracks are identified as kaons or pions by the number 
of photoelectrons detected in the ACC, the specific ionization energy 
loss $(dE/dx)$ in the CDC, and, if slow enough, their time-of-flight.
This information is used to calculate kaon and pion relative likelihoods 
${\cal L}^{}_K$ and  ${\cal L}^{}_\pi$.
Tracks are identified as pions or kaons based on the likelihood ratios
$R^{}_{K,\pi}\equiv{\cal L}^{}_{K,\pi}/({\cal L}^{}_\pi + {\cal L}^{}_K)$.
The efficiency for kaons is typically 84\% with a pion 
misidentification rate of 5\%; the efficiency for pions
is typically 91\% with a kaon misidentification rate of 10\%.

For most final states there is substantial background from 
$e^+e^-\rightarrow q\bar{q}$ continuum events ($q=u,d,s,c$).
We distinguish this background from $B$ decays by first combining 
five modified Fox-Wolfram moments into a Fisher discriminant. 
This is then combined with a likelihood function for $\theta^{}_B$, 
the polar angle of the $B$ meson flight direction, and the resulting 
likelihood function is used to form a likelihood ratio
$R^{}_{q\bar{q}}\equiv{\cal L}^{}_{sig}/({\cal L}^{}_{sig} + {\cal L}^{}_{q\bar{q}})$.
A mode-dependent cut on \Rqq\ is made to significantly
reduce background events. Typical cut values reject $\simge 90$\% 
of \qqbar\ background with a signal efficiency of 40--70\%.

The analyses presented here proceed in three steps:
{\it (a)}\ selecting the final state of interest using 
$R^{}_{K,\pi}$ to identify tracks as pions or kaons;
{\it (b)}\ using \Rqq\ to reject continuum background; and
{\it (c)}\ selecting $B$ decays by cutting on the variables
$m^{}_{bc} \equiv \sqrt{E^{*\,2}_{\rm beam} - p^{*\,2}_B}$ and
$\Delta E\equiv E^*_B - E^*_{\rm beam}$, where $E^*_{\rm beam}$ 
denotes the beam energy and $p^*_B$ and $E^*_B$ denote the 
reconstructed momentum and energy of the candidate $B$ meson, 
all evaluated in the $e^+e^-$ CM frame. For 
correctly-identified $B$ decays, 
$m^{}_{bc} = M^{}_B$ and $\Delta E = 0$.
Throughout this paper, charge-conjugate modes are included unless
stated otherwise. When two errors are listed for a measurement,
the first one is statistical and the second one systematic.

\section{$B\rightarrow\pi\pi/K\pi/KK$ decays}

\begin{figure}
\vbox{
\hbox{
  \includegraphics[height=.20\textheight]{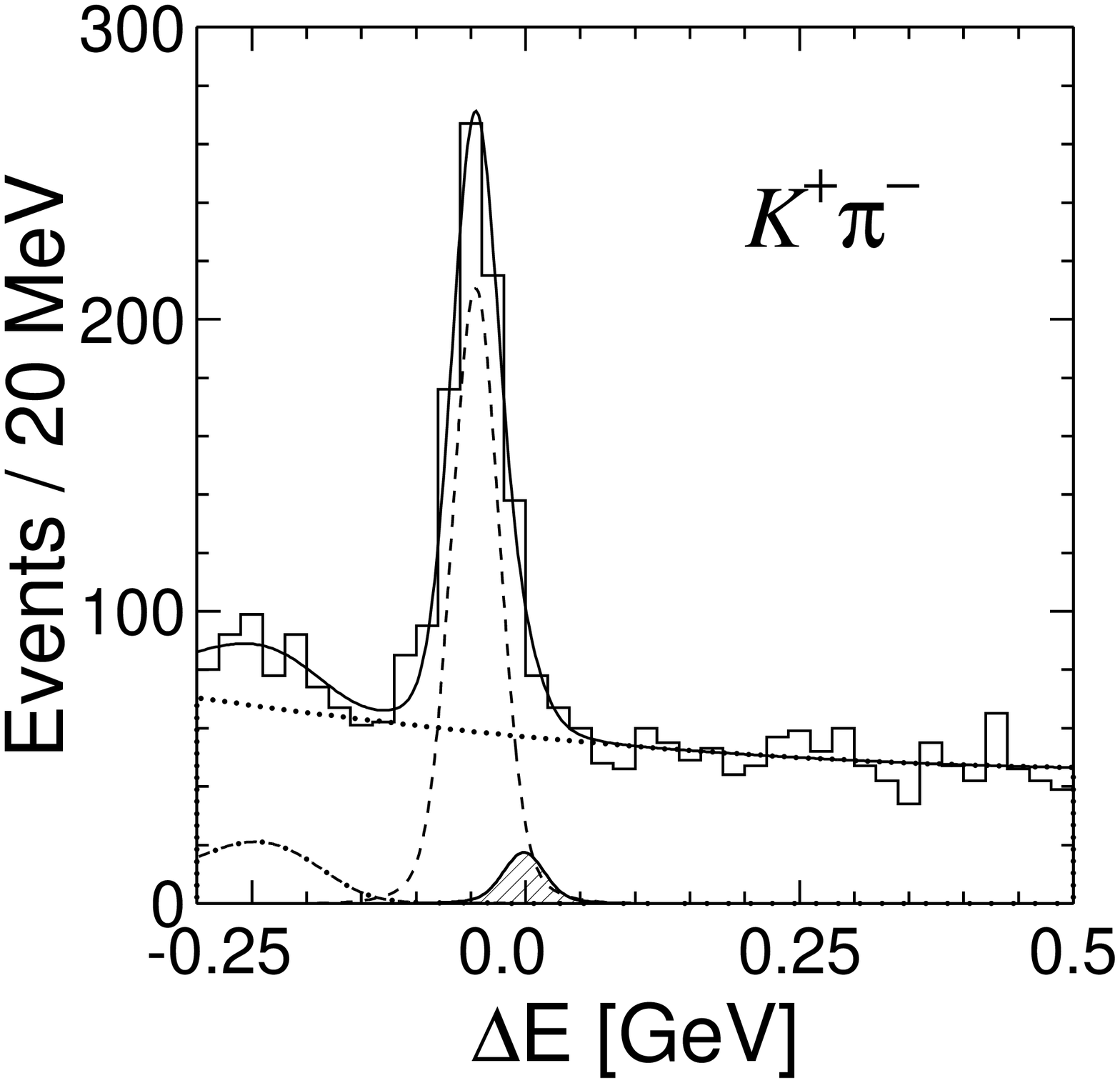}
\hskip0.20in
  \includegraphics[height=.20\textheight]{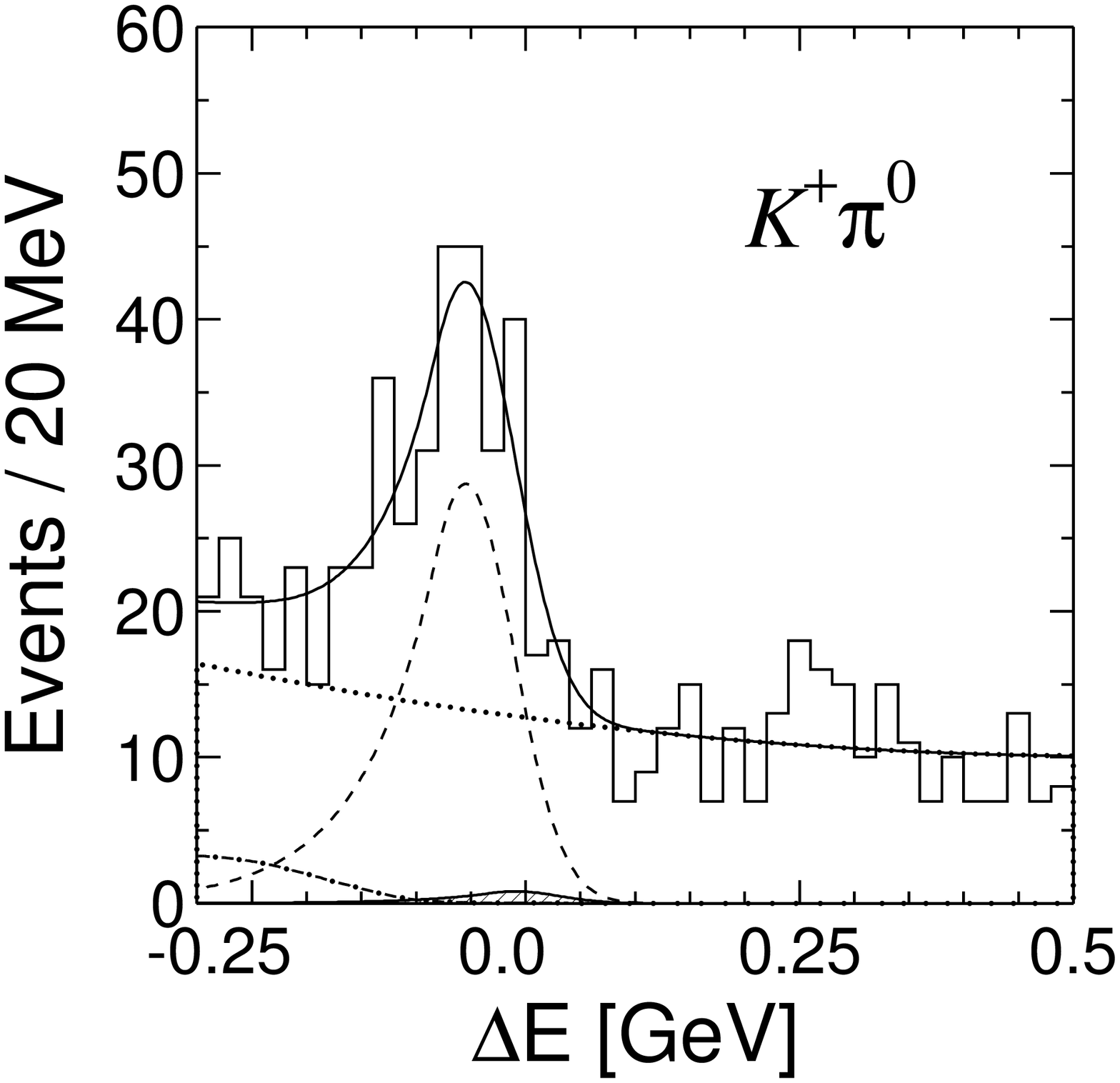}
\hskip0.20in
  \includegraphics[height=.20\textheight]{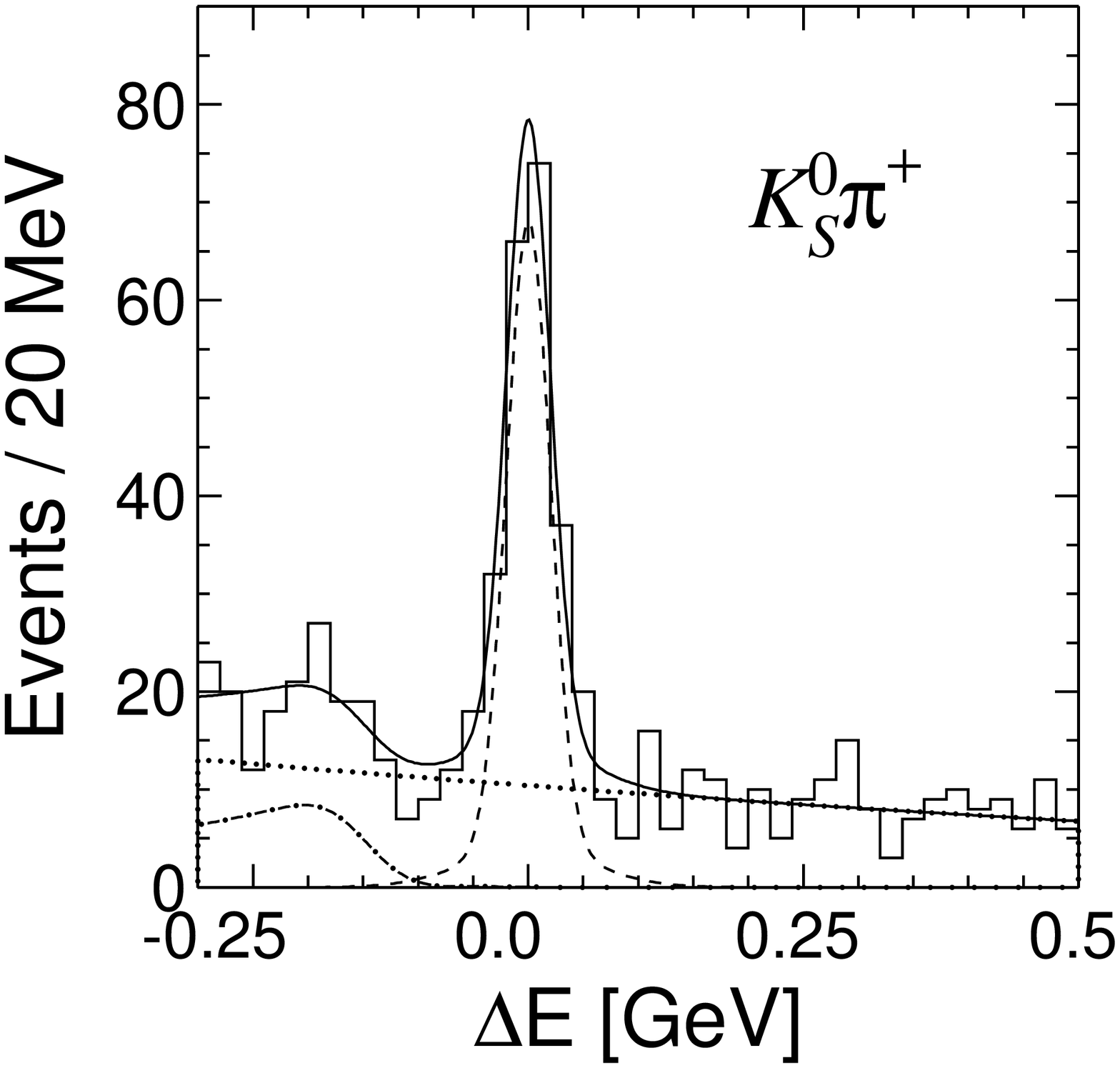}
}
\vskip0.02in
\hbox{
  \includegraphics[height=.20\textheight]{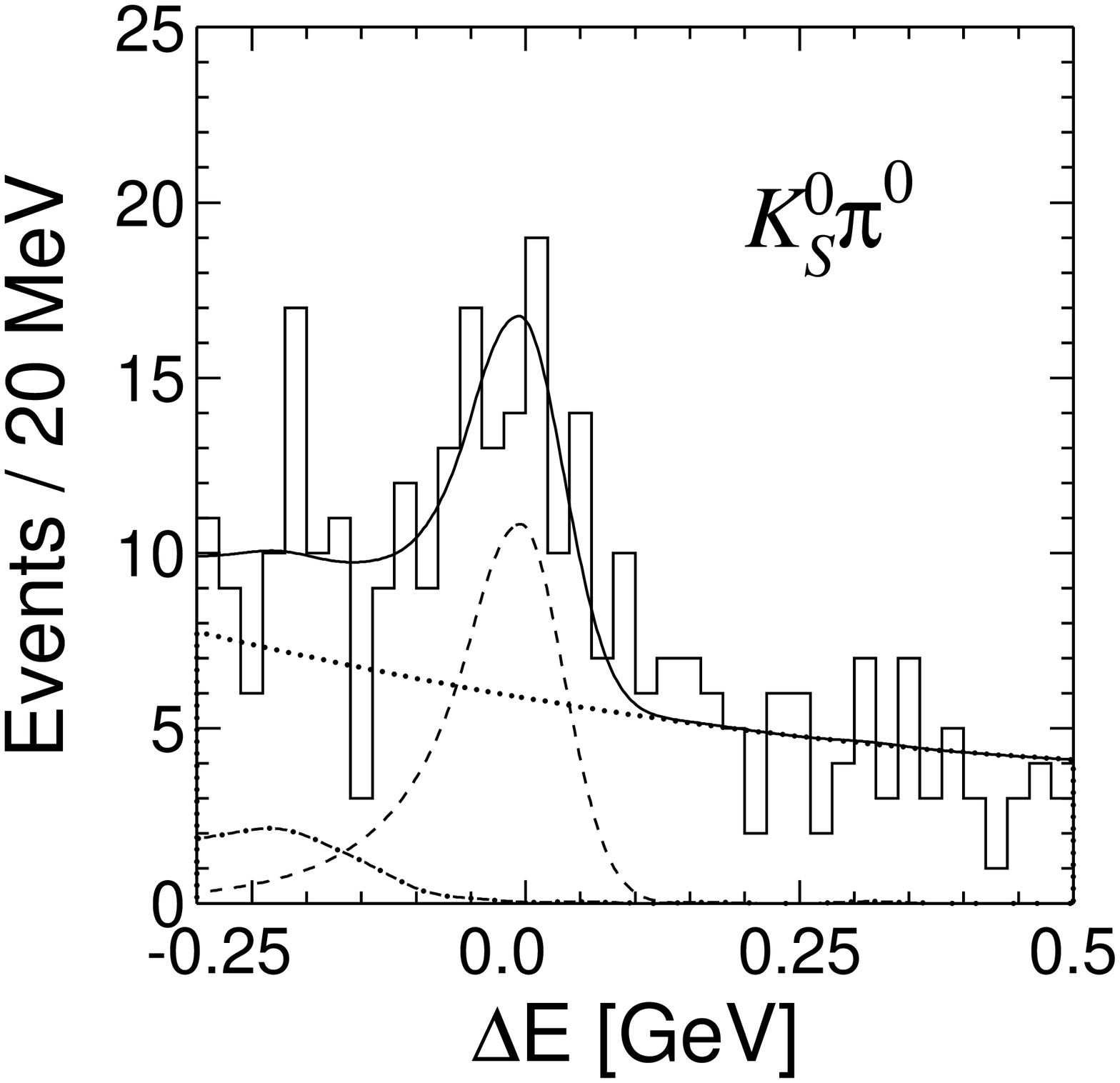}
\hskip0.20in
  \includegraphics[height=.20\textheight]{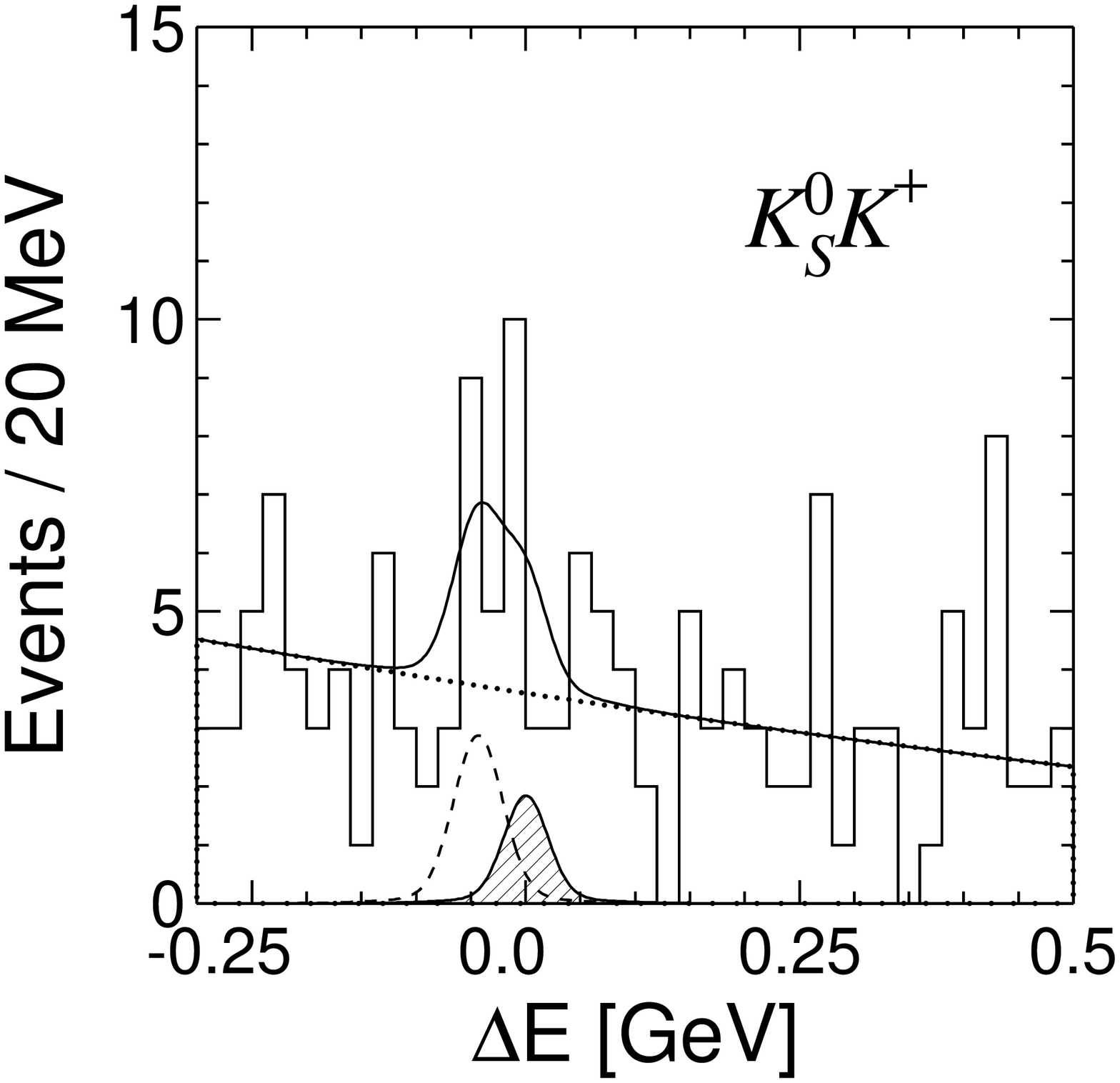}
\hskip0.20in
  \includegraphics[height=.20\textheight]{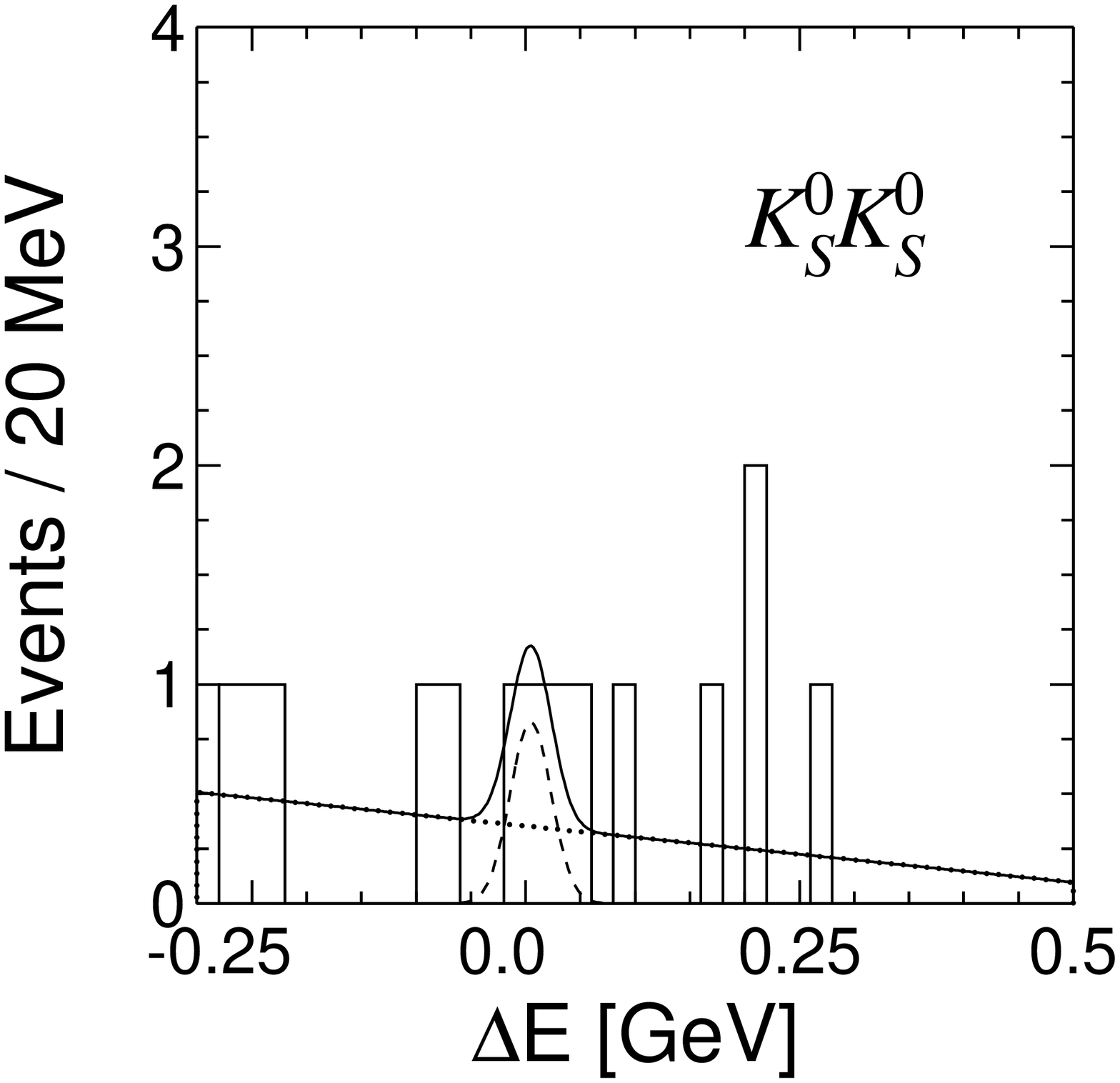}
}
\vskip0.02in
\hbox{
  \includegraphics[height=.20\textheight]{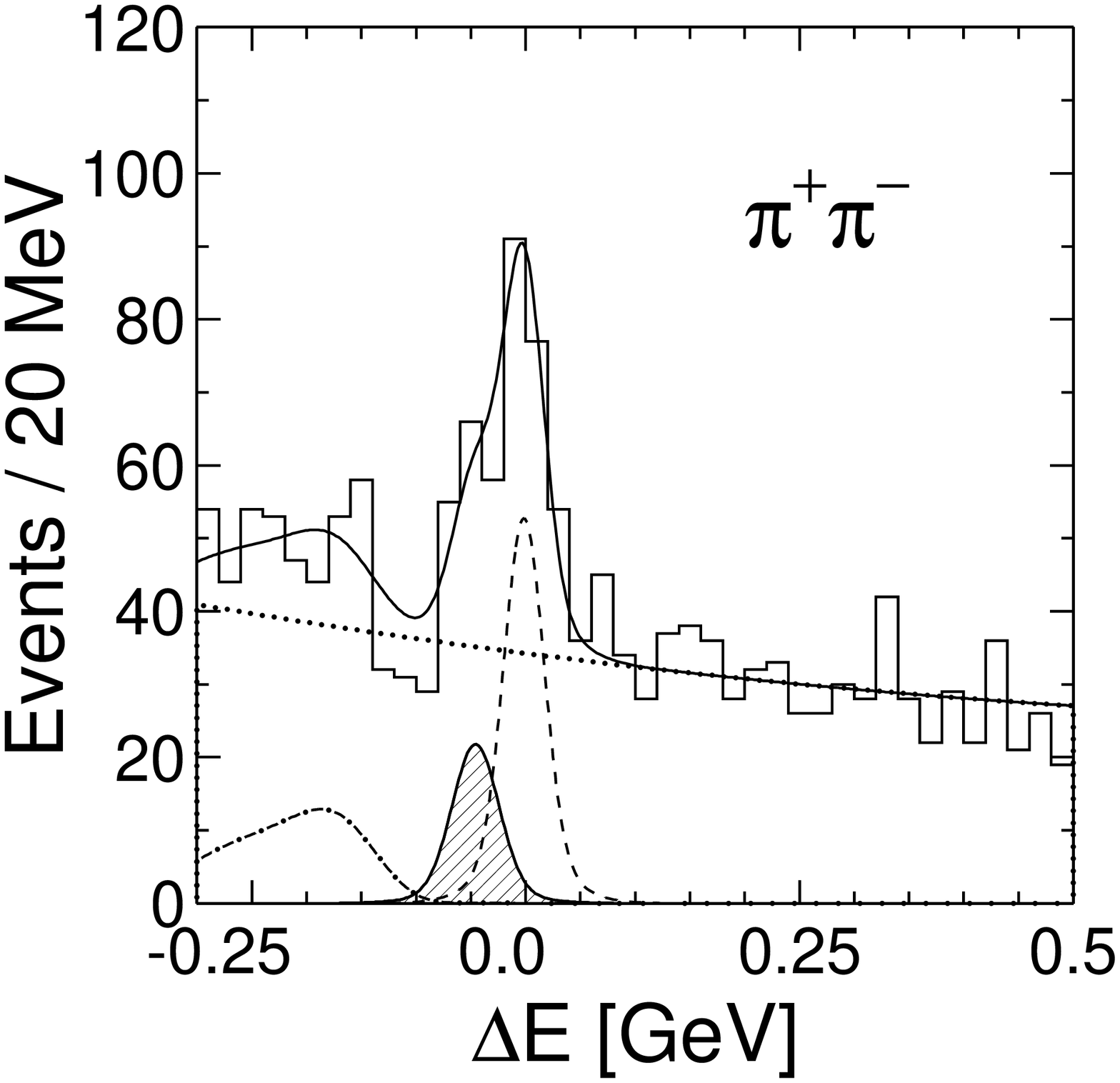}
\hskip0.20in
  \includegraphics[height=.20\textheight]{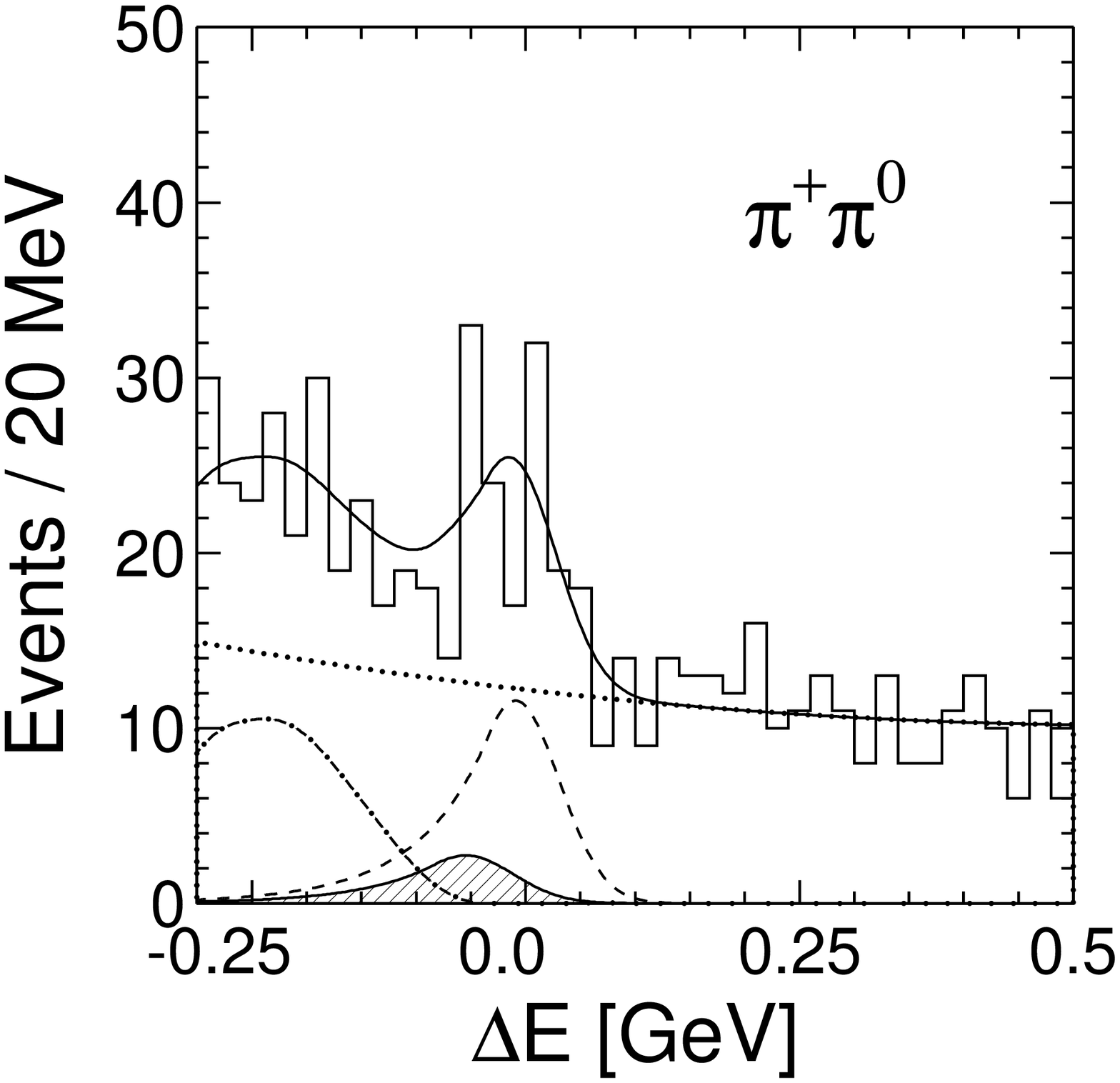}
\hskip0.20in
  \includegraphics[height=.20\textheight]{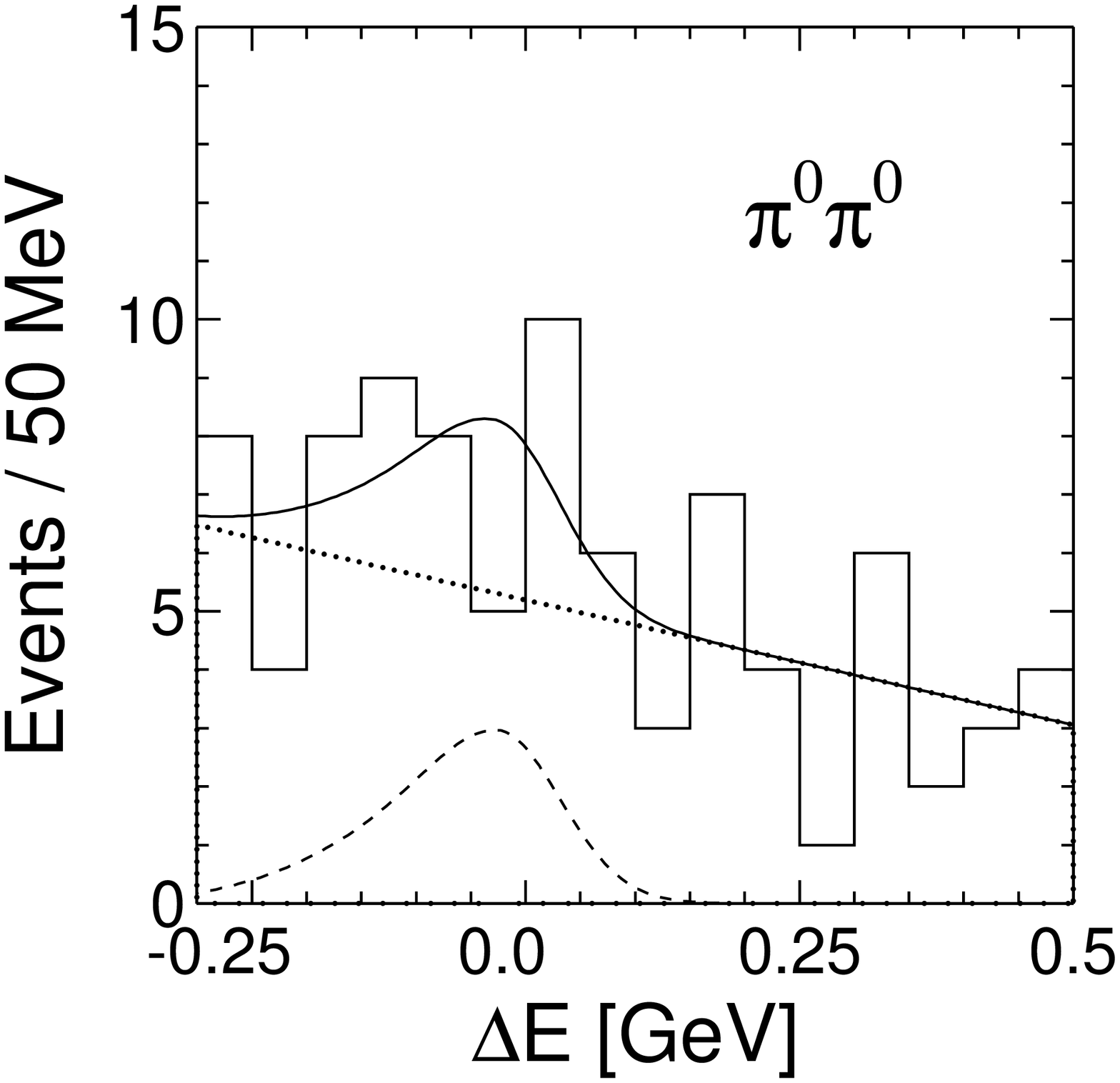}
}
}
\caption{\bhh\ $\Delta E$ distributions for 
$5.27<m^{}_{bc}<5.29$\gevm. The fit results are shown as the solid, 
dashed, dotted, and dash-dotted curves for the total, signal, \qqbar\ 
background, and $B\bar{B}$ background, respectively. The hatched area 
indicates reflections resulting from $\pi^\pm \rightarrow K^\pm$ 
misidentification. All tracks are assigned the pion mass; this produces 
the shift of signal modes containing $K^\pm$'s towards $-\Delta E$ values.
\label{fig:hh_deltaE}}
\end{figure}

These decays proceed via $b\rightarrow u$ tree and $b\rightarrow d,s$ loop
diagrams, and the final states include both charged and neutral kaons and pions.
Neutral kaons are identified via $K^0_S\rightarrow\pi^+\pi^-$, and neutral
pions are identified via $\pi^0\rightarrow\gamma\gamma$ (where the photons 
produce clusters in the ECL). The $\Delta E$ distributions after all 
selection cuts and a cut $5.27<m^{}_{bc}<5.29$\gevm\ are shown in 
Fig.~\ref{fig:hh_deltaE}. The event yields are obtained by fitting 
these distributions for signal, \qqbar\ 
background, and other charmless $B$ decay background. Possible
reflections due to $K^\pm/\pi^\pm$ misidentification are included where
applicable. All fit parameters other than the normalizations are fixed:
most are determined from Monte Carlo (MC) simulation, while others
are determined directly from the data, usually from events in a
lower $m^{}_{bc}$ sideband. The resulting event yields are
listed in Table~\ref{tab:hh_br}. The statistical significance of 
signals is calculated as ${\cal S} = \sqrt{2\ln(L^{}_{max}/L^{}_0)}$,
where $L^{}_0$ is the likelihood obtained assuming no signal events and
$L^{}_{max}$ is the (maximum) likelihood obtained with $N^{}_s$ signal
events. For cases where no significant signal is observed, we quote a 
90\% C.L. upper limit using a Feldman-Cousins frequentist approach~\cite{FeldCous}.
The corresponding branching fractions are also listed in Table~\ref{tab:hh_br}
and show the theoretically-expected hierarchy: 
$B(B\rightarrow K\pi)>B(B\rightarrow\pi\pi)>B(B\rightarrow K\bar{K})$.

\begin{table}
\renewcommand{\arraystretch}{1.2}
\begin{tabular}{l|crrl}
\hline
\tablehead{1}{c}{b}{Mode} & 
\tablehead{1}{c}{b}{$N^{}_s$} & 
\tablehead{1}{c}{b}{${\cal S}$} & 
\tablehead{1}{c}{b}{$\varepsilon\ (\%)$} & 
\tablehead{1}{c}{b}{$B\!\times\! 10^{6}$ (90\% C.L.\,limit)} \\ 
\hline
{\boldmath $\pi^+\pi^-$} & $133\,^{+19}_{-18}$ & 8.5 & 35.2 & $\ \,4.4\,\pm\,0.6\,\pm\,0.3$ \\
{\boldmath $\pi^+\pi^0$} & $72.4\,\pm\,17.4$ & 4.5 & 16.1 & $\ \,5.3\,\pm\,1.3\,\pm\,0.5$ \\
{\boldmath $\pi^0\pi^0$} & $12.0\,^{+9.1}_{-8.6}$ & 1.9 & 7.8 & $\ \,1.8\,^{+1.4}_{-1.3}\,^{+0.5}_{-0.7}\ \ (<4.4)$ \\
\hline
{\boldmath $K^+\pi^-$} & $596\,\pm\,33$ & 24.1 & 37.9 & $18.5\,\pm\,1.0\,\pm\,0.7$ \\
{\boldmath $K^+\pi^0$} & $199\,\pm\,22$ & 10.8 & 18.3 & $12.8\,\pm\,1.4\,^{+1.4}_{-1.0}$ \\
{\boldmath $K^0\pi^+$} & $187\,\pm\,16$ & 16.4 & 10.0 & $22.0\,\pm\,1.9\,\pm\,1.1$ \\
{\boldmath $K^0\pi^0$} & $72.6\,\pm\,14.0$ & 5.8 & 6.8 & $12.6\,\pm\,2.4\,\pm\,1.4$ \\
\hline
{\boldmath $K^+ K^-$} & $-1.0\,^{+6.6}_{-5.9}$ & $-$ & 20.1 & $\!\!<\!0.7$ \\
{\boldmath $K^+\overline{K^0}$} & $8.6\,\pm\,5.9$ & 1.6 & 5.9 & $\ \,1.7\,\pm\,1.2\,\pm\,0.1\ \ (<3.4)$ \\
{\boldmath $K^0\overline{K^0}$} & $2.0\,\pm\,1.9$ & 1.3 & 2.9 & $\ \,0.8\,\pm\,0.8\,\pm\,0.1\ \ (<3.2)$ \\
\hline
\end{tabular}
\caption{Event yields, signal significance, efficiencies, and branching 
fractions (90\% C.L. upper limits) for $B\rightarrow hh$ decays.
\label{tab:hh_br}}
\end{table}

The branching fractions can be used to constrain the magnitudes of the 
CKM phases $\phi^{}_2$ and $\phi^{}_3$~\cite{phi2phi3constraints}. 
For such constraints, ratios of partial widths are most useful 
because of their reduced hadronic uncertainties. We thus use 
the results in Table~\ref{tab:hh_br} and the lifetime ratio 
$\tau^{}_{B^+}/\tau^{}_{B^0} = 1.083\,\pm\,0.017$~\cite{PDG} to
calculate the partial width ratios listed in Table~\ref{tab:hh_ratios}.
The fact that $\Gamma(\pi^+\pi^-)/2\Gamma(\pi^+\pi^0)\neq 1$ implies
that the penguin contribution to \bpipi\ is significant.

\begin{table}
\renewcommand{\arraystretch}{1.2}
\begin{tabular}{c|c}
\hline
\tablehead{1}{c}{b}{Ratio} & 
\tablehead{1}{c}{b}{Measured Value} \\
\hline
{\boldmath $\Gamma(\pi^+\pi^-) / 2\Gamma(\pi^+\pi^0)$} & 
				$0.45\,\pm\,0.13\,\pm\,0.05$ \\
{\boldmath $\Gamma(\pi^+\pi^-) / \Gamma(K^+\pi^-) $} & 
				$0.24\,\pm\,0.04\,\pm\,0.02$ \\
{\boldmath $\Gamma(\pi^0\pi^0) / \Gamma(\pi^+\pi^0)$} & 
				$< 0.92$\ \,@\,90\% C.L. \\
\hline
{\boldmath $2\Gamma(K^+\pi^0) / \Gamma(K^0\pi^+)$} & 
				$1.16\,\pm\,0.16\,^{+0.14}_{-0.11}$ \\
{\boldmath $\Gamma(K^+\pi^-) / \Gamma(K^0\pi^+)$} & 
				$0.91\,\pm\,0.09\,\pm\,0.06$ \\
{\boldmath $\Gamma(K^+\pi^-) / 2\Gamma(K^0\pi^0)$} & 
				$0.74\,\pm\,0.15\,\pm\,0.09$ \\
\hline
\end{tabular}
\caption{Partial width ratios for \bhh.
\label{tab:hh_ratios}}
\end{table}

For the flavor-specific decays $B\rightarrow K^\pm\pi^\mp,\,K^\pm\pi^0$, 
$K^0\pi^\pm$, and $\pi^\pm\pi^0$, the $\Delta E$ distributions are fitted 
separately for $B$ and $\overline{B}$ candidates to measure the \cp\ asymmetry
$A^{}_{CP}\equiv \left[N(\overline{B}\rightarrow\bar{f}) - N(B\rightarrow f)\right]/ 
\left[N(\overline{B}\rightarrow\bar{f}) + N(B\rightarrow f)\right]$,
where $B\,(\overline{B})$ represents $B^0$ or $B^+$ ($\overline{B^0}$ or $B^-$).
The results are listed in Table~\ref{tab:hh_asymmetries}; 
no significant \cp\ asymmetries are observed.

\begin{table}
\renewcommand{\arraystretch}{1.5}
\begin{tabular}{l|lllc}
\hline
\tablehead{1}{c}{b}{Mode} & 
\tablehead{1}{c}{b}{$N^{}_s(\overline{B})$} & 
\tablehead{1}{c}{b}{$N^{}_s(B)$} & 
\tablehead{1}{c}{b}{$A^{}_{CP}$} & 
\tablehead{1}{c}{b}{90\% C.L.\ Interval} \\ 
\hline
{\boldmath $\pi^+\pi^0$} & 	$\ \,31.2\,\pm\,11.9$ & $\ \,41.3\,\pm\,12.7$ & 
		$-0.14\,\pm\,0.24\,^{+0.05}_{-0.04}$ & $( -0.57,\ 0.30 )$  \\ 
\hline
{\boldmath $K^+\pi^-$} & 	$235.4\,^{+19.8}_{-19.1}$ & $270.2\,^{+19.7}_{-18.9}$ & 
		$-0.07\,\pm\,0.06\,\pm\,0.01$ & $( -0.18,\ 0.04 )$  \\ 
{\boldmath $K^+\pi^0$} & 	$122.0\,\pm\,15.8$ & $\ \,76.5\,\pm\,14.5$ & 
		$\ \ \,0.23\,\pm\,0.11\,^{+0.01}_{-0.04}$ & $( -0.01,\ 0.42 )$  \\ 
{\boldmath $K^0\pi^+$} & 	$119.1\,^{+13.8}_{-13.1}$ & $104.4\,^{+13.2}_{-12.5}$ & 
		$\ \ \,0.07\,^{+0.09}_{-0.08}\,^{+0.01}_{-0.03}$ & $( -0.10,\ 0.22 )$  \\ 
\hline
\end{tabular}
\caption{\cp\ asymmetries for $B\rightarrow hh$. For \bpipi$\!\!\!$, see~\cite{cp_pipi}
($t$-dependent analysis).
\label{tab:hh_asymmetries}}
\end{table}

\section{$B^\pm\rightarrow D^{}_{CP}K^\pm$ decays}

\begin{figure}
\includegraphics[height=0.15\textheight,width=.55\textheight]{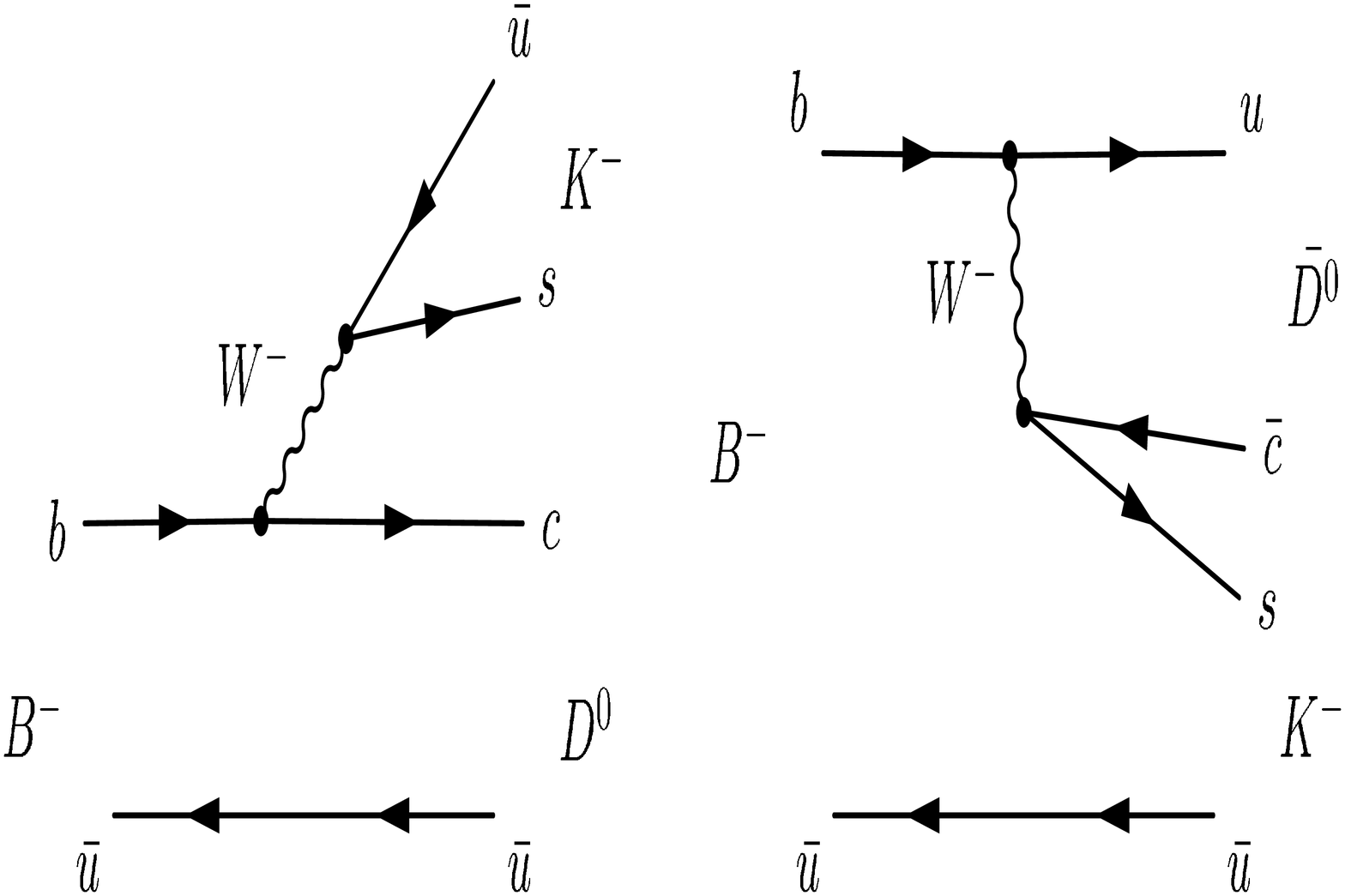}
\caption{Feynman diagrams for \bdcpk: $b\rightarrow c$ tree (left) and 
$b\rightarrow u$ tree (right).
\label{fig:bdcpk_diagrams}}
\end{figure}

The decay \bdcpk, where \dcp\ represents a $D^0$ decaying to 
a \cp\ eigenstate, proceeds via $b\rightarrow c$ and $b\rightarrow u$ 
transitions as shown in Fig.~\ref{fig:bdcpk_diagrams}. Interference between
the amplitudes gives rise to direct \cp\ violation, and measuring \acp\ 
allows one to constrain the CKM phase $\phi^{}_3$.
The observables are~\cite{formulae_dcpk}:
\begin{eqnarray}
A^{}_{1,2} & \equiv & 
	\frac{B(B^-\!\rightarrow\!D^{}_{1,2}K^-) - B(B^+\!\rightarrow\!D^{}_{1,2}K^+)}
	     {B(B^-\!\rightarrow\!D^{}_{1,2}K^-) + B(B^+\!\rightarrow\!D^{}_{1,2}K^+)} 
\ =\ \frac{2r\sin\delta'\,\sin\phi^{}_3}{1+r^2+2r\cos\delta'\,\cos\phi^{}_3} \nonumber \\
{\cal R}^{}_{1,2} & \equiv & \frac{R^{D^{1,2}}}{R^{D^0}} = 
				1+r^2 + 2r\cos\delta'\,\cos\phi^{}_3\,,    
\label{eqn:bdcpk_formulae}
\end{eqnarray}
where $\delta' = \delta\,(\delta+\pi)$ for $D^{}_1\,(D^{}_2)$
and the ratios $R^{D^{1,2}}$ and $R^{D^0}$ are:
\begin{eqnarray*}
R^{D^{1,2}} & = & 
	\frac{B(B^-\!\rightarrow\!D^{}_{1,2}K^-) + B(B^+\!\rightarrow\!D^{}_{1,2}K^+)}
	     {B(B^-\!\rightarrow\!D^{}_{1,2}\pi^-) + B(B^+\!\rightarrow\!D^{}_{1,2}\pi^+)} \\
 & & \\
R^{D^0} & = & 
	\frac{B(B^-\!\rightarrow\!D^0 K^-) + B(B^+\!\rightarrow\!\overline{D^0}K^+)}
	     {B(B^-\!\rightarrow\!D^0 \pi^-) + B(B^+\!\rightarrow\!\overline{D^0}\pi^+)}\,.
\end{eqnarray*}
In these expressions, $D^{}_1$ and $D^{}_2$ are \cp-even and \cp-odd eigenstates,
respectively, of the neutral $D^0$ meson; $r$ is the ratio of the $b\rightarrow u$
and $b\rightarrow c$ amplitudes shown in Fig.~\ref{fig:bdcpk_diagrams}; and $\delta$
is their strong phase difference. The ratio $r$ is expected to be only $\sim$\,0.1
due to a CKM suppression factor and a color suppression factor. The ratio 
$R^{D^0}$ has been previously measured by 
\belle\ ($0.079\,\pm\,0.009\,\pm\,0.006$~\cite{rd0_belle})
and CLEO ($0.099\,^{+0.014}_{-0.012}\,^{+0.007}_{-0.006}$~\cite{rd0_cleo}).
The results are in agreement with naive factorization: 
$\tan^2\theta^{}_C\,(f^{}_K/f^{}_\pi)^2\approx 0.074$.
Here we present new results for $R^{D^0}\!\!\!$, $R^{D^{1,2}}\!\!\!$, and 
$A^{}_{1,2}$; these values can be inserted into Eq.~(\ref{eqn:bdcpk_formulae}) 
to determine the three unknowns $r$, $\delta$, and $\phi^{}_3$.

For this analysis 
$D^0$ mesons are reconstructed as $K^-\pi^+$;
$D^{}_1$ mesons ($CP=+1$) as $K^+ K^-$ and $\pi^+\pi^-$; and 
$D^{}_2$ mesons ($CP=-1$) as $K^0_S\pi^0$, $K^0_S\phi$, $K^0_S\omega$, 
$K^0_S\eta$, and $K^0_S\eta'$.
The short-lived mesons are reconstructed as follows:
$\phi\rightarrow K^+K^-$ with $1.008<m^{}_{KK}<1.032$\gevm;
$\omega\rightarrow \pi^+\pi^-\pi^0$ with $0.732<m^{}_{\pi\pi\pi}<0.820$\gevm;
$\eta\rightarrow\gamma\gamma$ with $0.495<m^{}_{\gamma\gamma}<0.578$\gevm; and
$\eta'\rightarrow\eta\pi^+\pi^-$ with $0.903<m^{}_{\eta\pi\pi}<1.002$\gevm.
The resulting $D^0$ candidates are required to have masses within $2.5\sigma$
of $m^{}_{D^0}$, where $\sigma$ is the measured mass resolution
($4.5\!-\!18$\mevm). The $D^0$ and $\pi^+/K^+$ 
candidates are combined to form $B^+$ candidates by selecting combinations
with $5.27<m^{}_{bc}<5.29$\gevm\ and $\left|\Delta E\right|<0.20$\geve.
The event yields are obtained from fits to the $\Delta E$ distributions. 
The results 
for $R^{D^0}\!\!\!$, $R^{D^{1,2}}\!\!\!$, and $A^{}_{1,2}$ are listed in 
Table~\ref{tab:bdcpk_results}; all \cp\ asymmetries are consistent with zero.
The factor $r$ can be calculated via ${\cal R}^{}_1 + {\cal R}^{}_2 = 2(1+r^2)$;
the result is $r^2=0.31\,\pm\,0.21$, which is only $1.5\sigma$ from zero.
Since, in Eq.~(\ref{eqn:bdcpk_formulae}), $\cos\phi^{}_3$ and $\sin\phi^{}_3$
are always multiplied by a factor of $r$, the value of $r$ obtained precludes 
setting a stringent constraint upon $\phi^{}_3$ with the current statistics.
The situation should improve with more data.

\begin{table}
\renewcommand{\arraystretch}{1.2}
\begin{tabular}{l|cc}
\hline
\tablehead{1}{c}{b}{Mode} & 
\tablehead{1}{c}{b}{$A^{}_{CP}$} & 
\tablehead{1}{c}{b}{90\% C.L.\ Interval} \\
\hline
{\boldmath $D^0 K^\pm$} & $\ \ 0.04\,\pm\,0.06\,\pm\,0.03$ & $(-0.07,\ 0.15)$ \\
{\boldmath $D^{}_1 K^\pm$} & 	$\ \ 0.06\,\pm\,0.19\,\pm\,0.04$ & $(-0.26,\ 0.38)$ \\
{\boldmath $D^{}_2 K^\pm$} & 	$-0.19\,\pm\,0.17\,\pm\,0.05$ & $(-0.47,\ 0.11)$ \\
\hline
$R^{D^0}$   & $0.077\,\pm\,0.005\,\pm\,0.006$ & \\
$R^{D^{1}}$ & $0.093\,\pm\,0.018\,\pm\,0.008$ & \\
$R^{D^{2}}$ & $0.108\,\pm\,0.019\,\pm\,0.007$ & \\
\hline
\end{tabular}
\caption{Results for \acp\ (top) and
$R^{D^0}\!\!\!$, $R^{D^{1,2}}$ (bottom).
\label{tab:bdcpk_results}}
\end{table}

\section{$B\rightarrow\omega K/\omega\pi$ decays}

The decays \bomegak\ and \bomegapi\ also proceed via $b\!\rightarrow\!u$ 
tree and $b\!\rightarrow\!s$ loop diagrams. 
Theoretical calculations based on QCD factorization~\cite{qcdfact1,qcdfact2,qcdfact3,qcdfact4} 
predict $B(B\rightarrow\omega\pi)\approx 2\times B(B\rightarrow\omega K)$. 
A previous Belle measurement~\cite{omegaK_belle} based on 
29~fb$^{-1}$ of data did not agree with this prediction, 
and we update that result here.

Candidate events are selected by first selecting 
$\omega\!\rightarrow\!\pi^+\pi^-\pi^0$ decays. The $\pi^0$ is
reconstructed from $\gamma\gamma$ pairs having 
$| m^{}_{\gamma\gamma}-m^{}_{\pi^0} |<3\sigma$
($\sigma\!=\!5.4$\mevm); each $\gamma$ must also 
satisfy $E^{}_\gamma>50$\meve. We then require
$\left|m^{}_{\pi\pi\pi}-m^{}_\omega\right|<30$\mevm\ ($2\sigma$),
and the $\omega$ is paired with a $\pi^\pm$, $\pi^0$, $K^\pm$, 
or $K^0_S$ to form $B$ candidates.
Those candidates satisfying $5.20<m^{}_{bc}<5.30$\gevm\ and
$\left|\Delta E\right|<0.25$\geve\ are subjected to an 
unbinned maximum likelihood (ML) fit using $m^{}_{bc}$ and 
$\Delta E$ as the independent variables.
The event yields resulting from the fit and the corresponding
branching fractions are listed in Table~\ref{tab:omega_yields}. 
We note that the central value for $B(B\rightarrow\omega K)$ 
is still greater than that for $B(B\rightarrow\omega\pi)$,
in contrast with the theoretical prediction.

\begin{figure}
\includegraphics[height=.26\textheight]{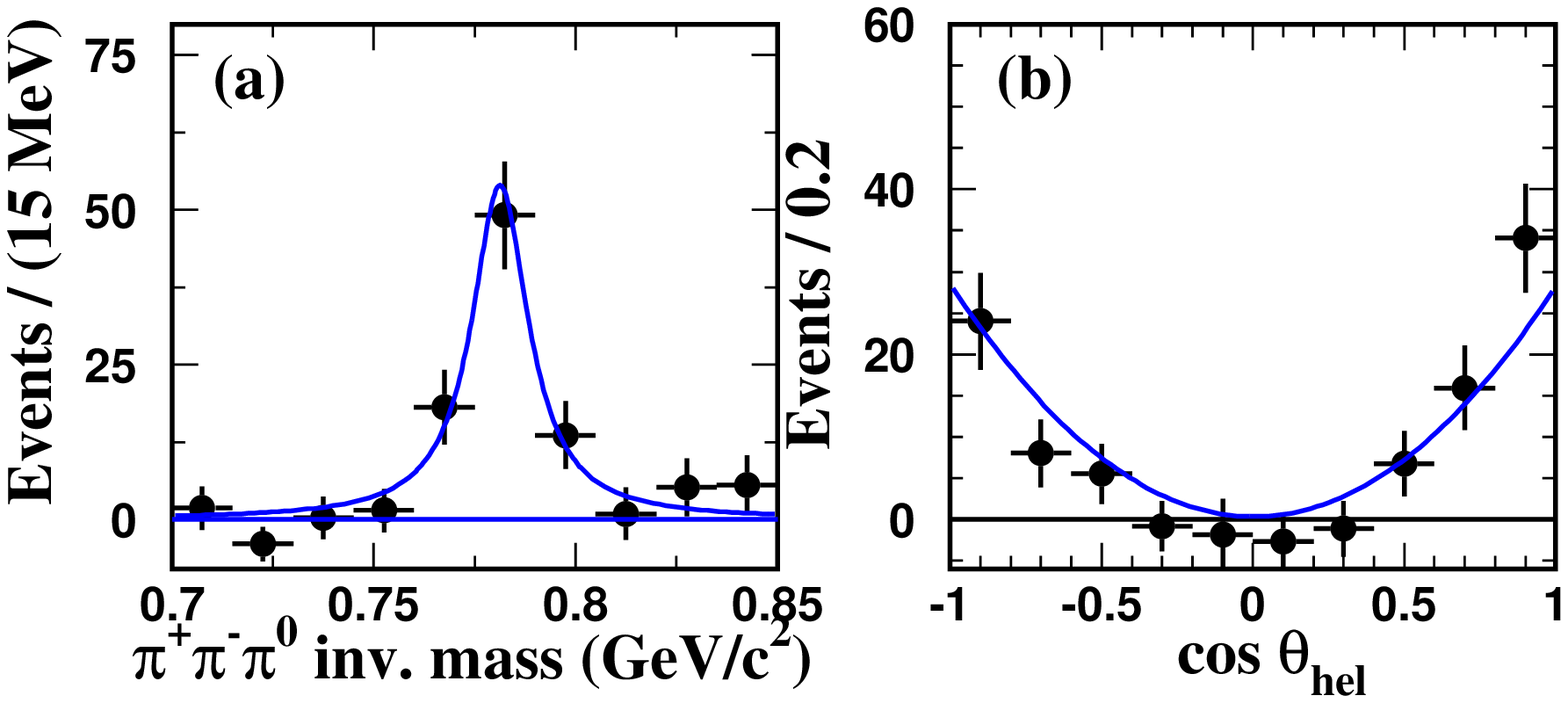}
\caption{The fitted event yields in bins of $m^{}_{\pi^+\pi^-\pi^0}$ (left)
and $\cos\theta^{}_h$ (right) for \bomegak\ and \bomegapi.
\label{fig:omega_mass}}
\end{figure}

The main background is due to \qqbar\ continuum events. 
To reduce this we cut on both \Rqq\ and the helicity
angle $\theta^{}_h$, which is defined as the angle between the $B$
flight direction and the vector perpendicular to the
$\omega$ decay plane, in the $\omega$ rest frame. The cut
chosen is $\left|\cos\theta^{}_h\right|>0.5$. To confirm
that signal candidates contain real $\omega$ decays, we 
relax the $m^{}_{\pi^+\pi^-\pi^0}$ cut and repeat the fits 
for different $m^{}_{\pi^+\pi^-\pi^0}$ bins. The resulting
event yields are plotted in Fig.~\ref{fig:omega_mass} and 
display a sharp peak at $m^{}_\omega$ with negligible 
nonresonant background underneath.

\begin{table}
\renewcommand{\arraystretch}{1.6}
\begin{tabular}{l|ccl}
\hline
\tablehead{1}{c}{b}{Mode} & 
\tablehead{1}{c}{b}{$N^{}_s$} & 
\tablehead{1}{c}{b}{${\cal S}$} & 
\tablehead{1}{c}{b}{$B\!\times\!10^6$ (90\% C.L.\ limit)} \\
\hline
{\boldmath $\omega K^-$} & $46.1\,^{+9.1}_{-8.4}$ & 7.8 & $\ \,6.7\,^{+1.3}_{-1.2}\,\pm\,0.6$ \\
{\boldmath $\omega K^0$} & 	$11.1\,^{+5.2}_{-4.4}$ & 3.2 & 
$\ \,4.0\,^{+1.9}_{-1.6}\,\pm\,0.5\ (< 7.6)$ \\
{\boldmath $\omega\pi^-$} & $42.1\,^{+10.1}_{-9.3}$ & 6.0 & $\ \,5.7\,^{+1.4}_{-1.3}\,\pm\,0.6$ \\
{\boldmath $\omega\pi^0$} & $0.0\,^{+2.1}_{-0.0}$ & $-$ & $(< 1.9)$ \\
\hline
\end{tabular}
\caption{\bomegak\ and \bomegapi\ event yields, statistical significance, 
and branching fractions (90\% C.L.\ limits).
\label{tab:omega_yields}}
\end{table}

Since the $B^\pm\rightarrow\omega h^\pm$ final states are self-tagging,
we divide these samples into $B^+$ and $B^-$ decays and search for a 
\cp\ asymmetry. The quantity measured is
$A^{}_{CP}= \left[N(B^-) - N(B^+)\right]/\left[N(B^-) + N(B^+)\right]$.
The event yields are determined from a two-dimensional binned fit in the
$m^{}_{bc}$-$\Delta E$ plane. The results are listed in
Table~\ref{tab:omega_cp}. While \acp$(\omega K)$ is consistent 
with zero, \acp$(\omega \pi)$ is $2.4\sigma$ above zero, and 
a symmetric 90\% C.L.\ interval excludes $A^{}_{CP}=0$.

\begin{table}
\renewcommand{\arraystretch}{1.6}
\begin{tabular}{l|cclc}
\hline
\tablehead{1}{c}{b}{Mode} & 
\tablehead{1}{c}{b}{$N(B^-)$} & 
\tablehead{1}{c}{b}{$N(B^+)$} & 
\tablehead{1}{c}{b}{$A^{}_{CP}$} & 
\tablehead{1}{c}{b}{90\% C.L.\ Interval} \\
\hline
{\boldmath $\omega K^\pm$} & 	$24.3\,^{+6.7}_{-5.9}$ & $21.8\,^{+6.4}_{-5.7}$ & 
		$0.06\,^{+0.20}_{-0.18}\,\pm\,0.01$ & $(-0.25,\ 0.40)$ \\  
{\boldmath $\omega\pi^\pm$} & 	$32.5\,^{+8.2}_{-7.5}$ & $11.5\,^{+6.1}_{-5.3}$ & 
	$0.48\,^{+0.23}_{-0.20}\,\pm\,0.02$ & $(0.14,\ 0.86)$ \\  
\hline
\end{tabular}
\caption{$B^\pm\rightarrow\omega K^\pm$ and $B^\pm\rightarrow\omega\pi^\pm$ 
event yields separated by charge, and the resulting \cp\ asymmetry.
\label{tab:omega_cp}}
\end{table}

\section{$B\rightarrow\phi K/\phi K^*$ decays and polarization}

The decays \bphik\ and \bphikst\ proceed only via loop diagrams 
($b\!\rightarrow\!ss\bar{s}$) and thus are especially sensitive 
to new physics. Because both the $\phi$ and $K^*$ are spin-1, 
\bphikst\ is a mixture of \cp-even and \cp-odd states;  the 
individual components can be determined by measuring the 
$\phi$ polarization.

As a first step, \phikk\ decays are identified by requiring 
pairs of oppositely-charged tracks having $R^{}_K>0.1$ and 
$| m^{}_{KK}-m^{}_\phi |<10$\mevm. $K^*$ decays are 
reconstructed via $K^{*\,+}\rightarrow K^+\pi^0$, 
$K^{*\,+}\rightarrow K^0_S\pi^+$, and
$K^{*\,0}\rightarrow K^+\pi^-$; the resulting
two-body mass is required to be within 70\mevm\ of 
$m^{}_{K^*}$. \bphik\ (\bphikst) decays are 
selected by pairing a $\phi$ candidate with a $K$ ($K^*$) 
candidate and requiring that they be within the signal region
$5.271<m^{}_{bc}<5.289$ ($5.270\!-\!5.290$)\gevm\ 
and $\left|\Delta E\right|<0.64\,(0.60)$\geve.
The $\Delta E$ window is slightly larger for 
$K^{*\,+}\rightarrow K^+\pi^0$ decays due to 
shower leakage. 

The dominant background is due to \qqbar\ continuum events. 
There is also 5--9\% contamination of $B\rightarrow\phi K^{(*)}$ decays
from nonresonant $B\rightarrow K^+ K^- K^{(*)}$, and 2--12\% contamination 
from $B\rightarrow f(980)K^{(*)},\,f(980)\rightarrow K^+K^-$.
The uncertainty in the instrinsic width of the $f^{}_0(980)$
is included in the systematic error.

\begin{figure}
\hbox{
  \includegraphics[height=.32\textheight]{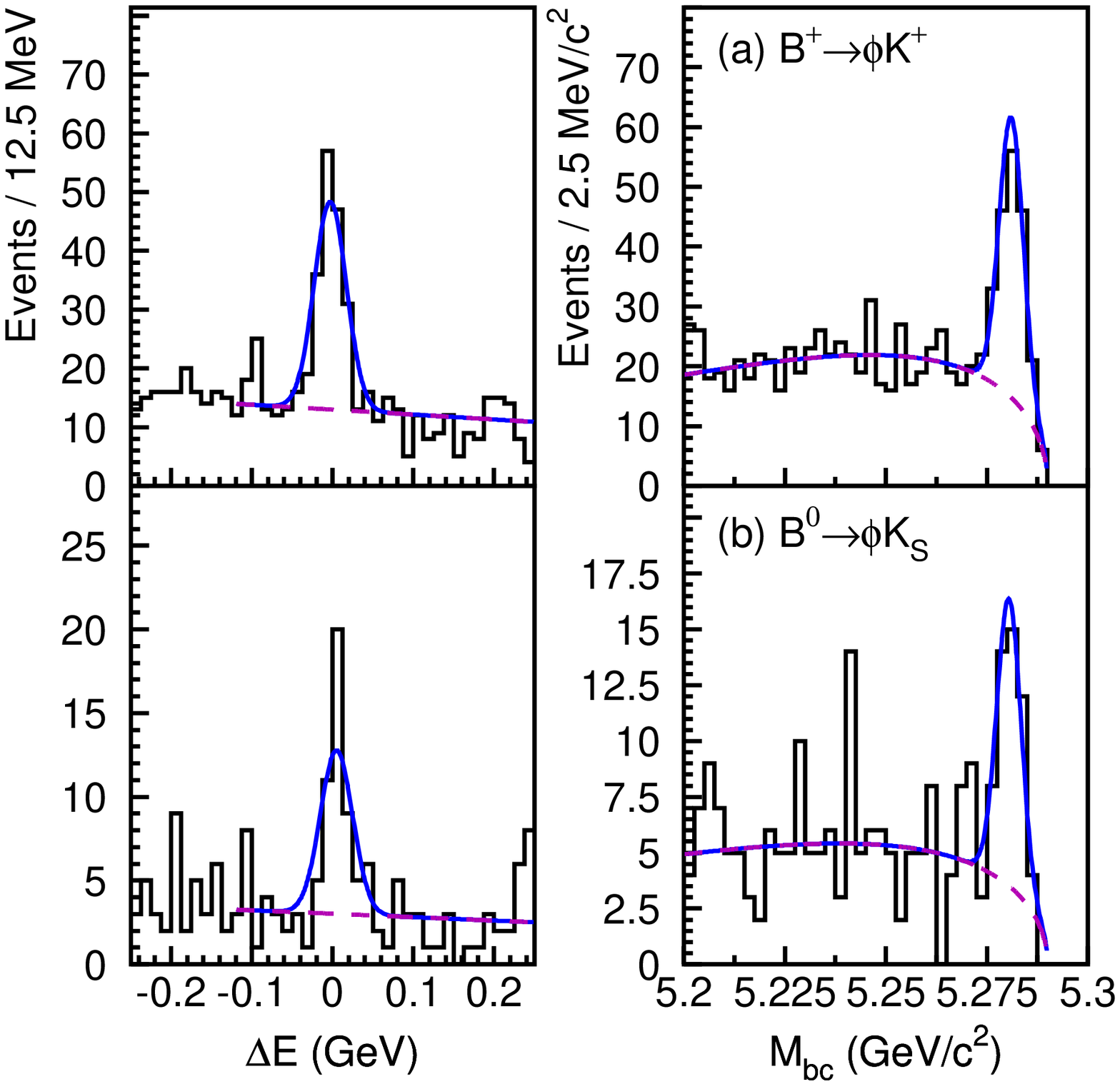}
\hskip0.30in
  \includegraphics[height=.32\textheight]{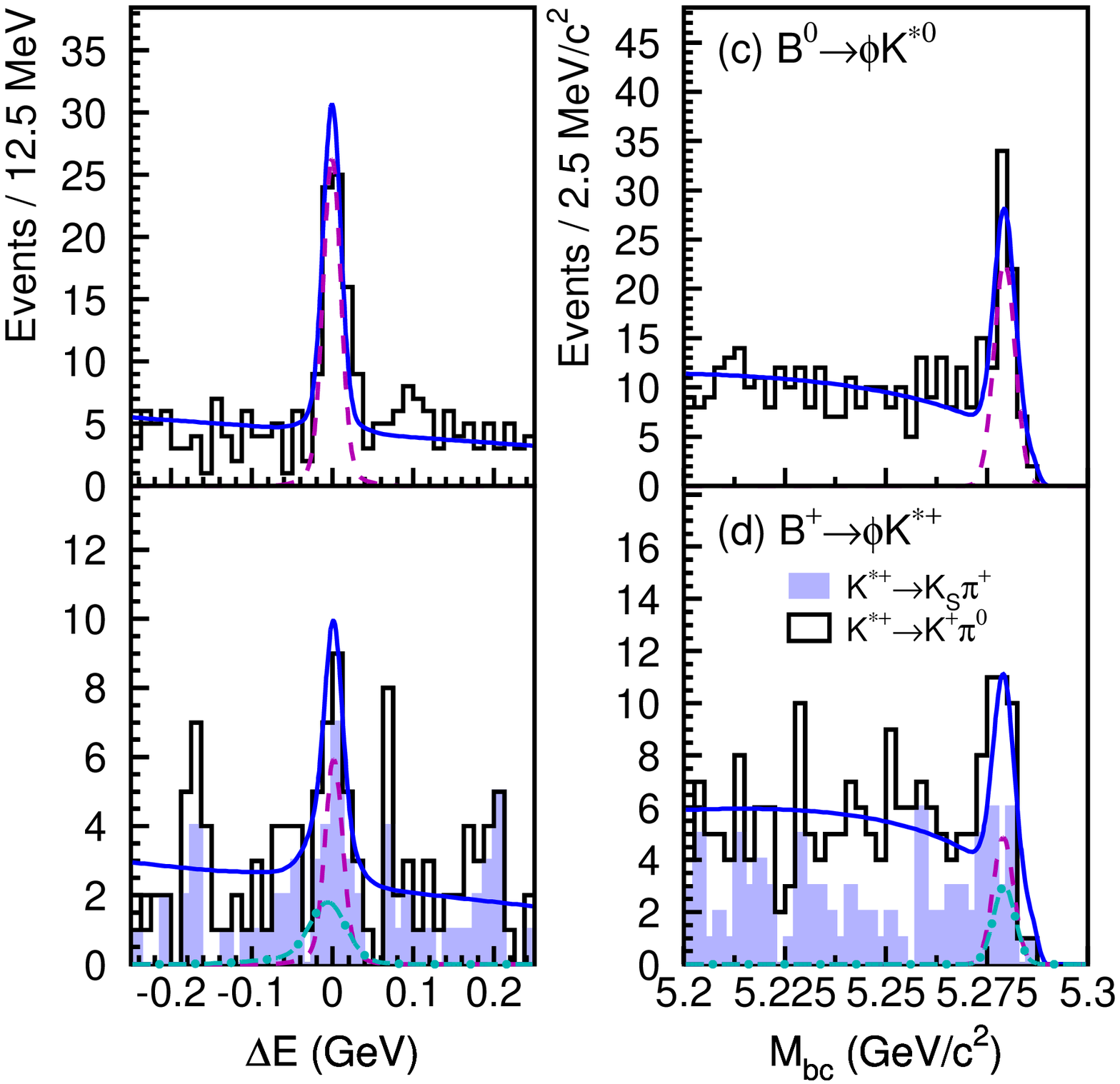}
}
\caption{Projections of the unbinned ML fits for \bphik\ decays (left)
and \bphikst\ decays (right). The histograms show the data. Events in 
the $m^{}_{bc}$ plots are required to have $|\Delta E |$ within the 
signal region, and events in the $\Delta E$ plots are required to 
have $m^{}_{bc}$ within the signal region (see text).
\label{fig:phikst_br}}
\end{figure}

The signal yields are obtained via an unbinned ML fit with
$m^{}_{bc}$ and $\Delta E$ as the independent variables.
The results and corresponding branching fractions are 
listed in Table~\ref{tab:phik_br}. The projections of 
the fits are shown in Fig.~\ref{fig:phikst_br}. For the 
\bphikst\ modes, there is an additional systematic error 
due to uncertainty in the $K^*$ polarization and the corresponding
uncertainty in the daughter $\pi$ detection efficiency.

\begin{table}
\renewcommand{\arraystretch}{1.6}
\begin{tabular}{l|clcc}
\hline
\tablehead{1}{c}{b}{Mode} & 
\tablehead{1}{c}{b}{$N^{}_s$} & 
\tablehead{1}{c}{b}{$B\times 10^6$} & 
\tablehead{1}{c}{b}{\acp} &
\tablehead{1}{c}{b}{90\% C.L.\ Interval} \\ 
\hline
{\boldmath $\phi K^+$} & $136\,^{+16}_{-15}$ &
	$9.4\,\pm\,1.1\,\pm\,0.7$ & $0.01\,\pm\,0.12\,\pm\,0.05$ & $(-0.20,\ 0.22)$ \\
{\boldmath $\phi K^0$} & $35.6\,^{+8.4}_{-7.4}$ &
			$9.0\,^{+2.2}_{-1.8}\,\pm\,0.7$ & $-$ & $-$ \\
{\boldmath $\phi K^{*\,0}$} & $58.5\,^{+9.1}_{-8.1}$ &
	$10.0\,^{+1.6}_{-1.5}{\,}^{+0.7}_{-1.8}$ & $0.07\,\pm\,0.15\,^{+0.05}_{-0.03}$ 
								& $(-0.18,\ 0.33)$ \\
{\boldmath $\phi K^{*\,+}$} & $\left\{
\begin{array}{ll}
8.0\,^{+4.3}_{-3.5}  & (K^+\pi^0) \\
11.3\,^{+4.5}_{-3.8} & (K^0_S\pi^+) \\
\end{array} \right.$ & 
$6.7\,^{+2.1}_{-1.9}{\,}^{+0.7}_{-1.0}$ & $-0.13\,\pm\,0.29\,^{+0.08}_{-0.11}$  
								& $(-0.64,\ 0.36)$ \\
\hline
\end{tabular}
\caption{Branching fractions and \cp\ asymmetries for \bphik\ and \bphikst\ decays.
\label{tab:phik_br}}
\end{table}

For the self-tagging modes $B^\pm\rightarrow\phi K^{(*)\,\pm}$
we measure $A^{}_{CP}= [N(\overline{B}) - N(B)]/[N(\overline{B}) + N(B)]$,
where $B\,(\overline{B})$ is $B^0$ or $B^+$ ($\overline{B^0}$ or $B^-$).
The results are also listed in Table~\ref{tab:phik_br} and
in all cases are consistent with zero.

\begin{figure}
\hbox{
\vbox{
  \includegraphics[height=.20\textheight]{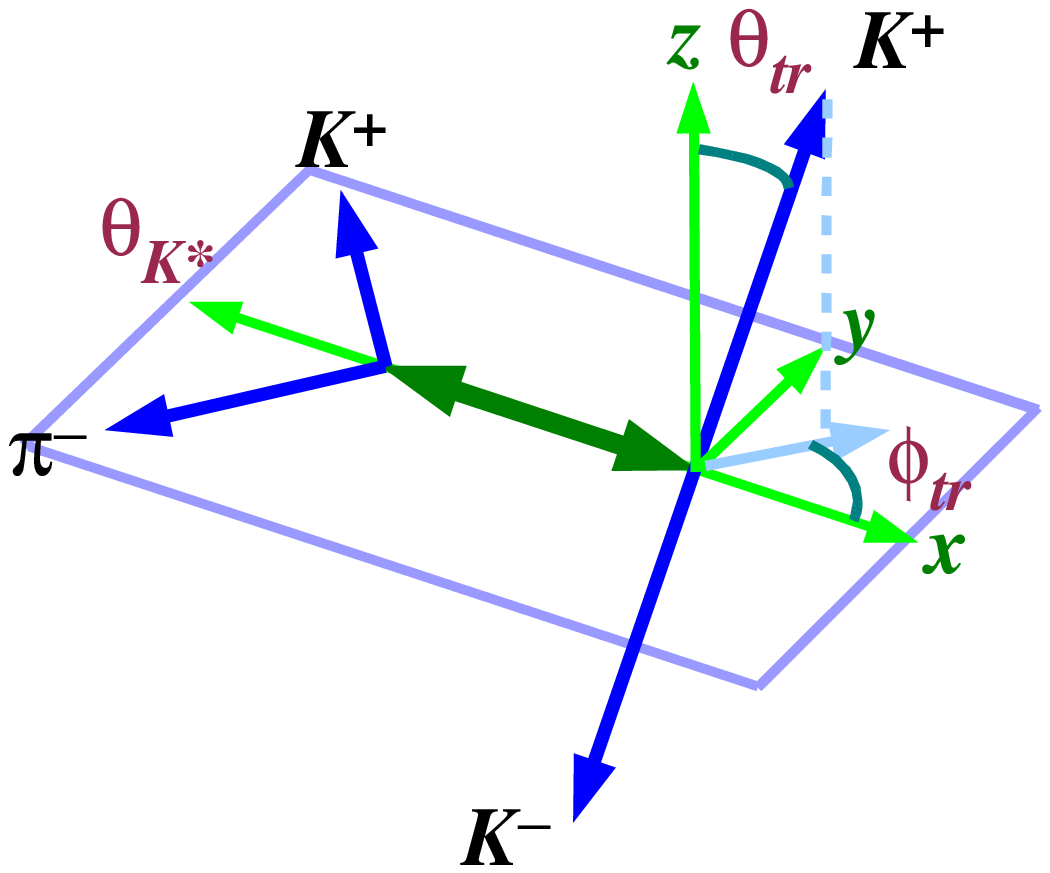}
\vskip0.8in
}
\hskip-3.5in
\includegraphics[height=.20\textheight]{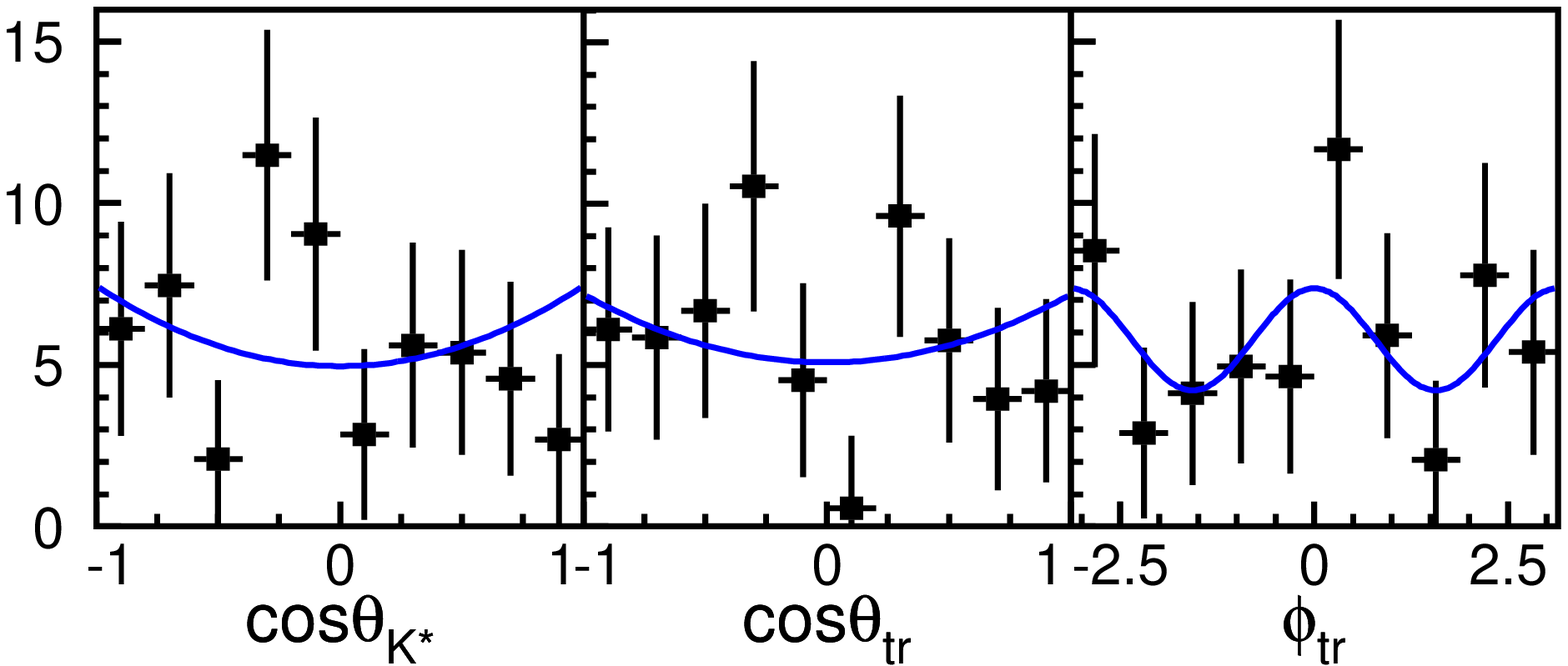}
}
\caption{Definition of the angles \thetakst, \thetatr, and 
\phitr\ in the transversity basis (left), and projections 
of the unbinned fit for these angles (right).
\label{fig:phikst_pol}}
\end{figure}

The polarization of the $\phi$ in \bphikst\ decays is measured using the 
transversity basis~\cite{transversity_basis}. In this basis the $\phi$ is 
at rest. The $x$-$y$ plane is defined by the $K^{*\,0}$ daughters, with 
the $-x$ axis along the direction of the $K^*$ (see Fig.~\ref{fig:phikst_pol}). 
The angle \thetakst\ is that between the $K^{*\,0}$ direction and the 
$K^+$ daughter. The angles \thetatr\ and \phitr\ are the 
polar and azimuthal angles, respectively, of the $K^+$ daughter of the 
$\phi$. The decay distribution is given by~\cite{triplediff}:
\begin{eqnarray}
\frac{d^3\Gamma(\phi^{}_{tr}, \cos\theta^{}_{tr}, \cos\theta^{}_{K^*})}
{d\phi^{}_{tr}\,d\cos\theta^{}_{tr}\,d\cos\theta^{}_{K^*}}
 & = &  \frac{9}{32\pi}\left[\,
	\left| A^{}_\perp\right|^2 2\cos^2\theta^{}_{tr}\sin^2\theta^{}_{K^*} \right.\nonumber\\
 & & \nonumber \\
 & \hskip-1.0in + & \hskip-0.5in 
  | A^{}_\parallel |^2 2\sin^2\theta^{}_{tr}\sin^2\phi^{}_{tr}\sin^2\theta^{}_{K^*} \nonumber\\
 & \hskip-0.8in + & \hskip-0.4in 
\left| A^{}_0\right|^2 4\sin^2\theta^{}_{tr}\cos^2\phi^{}_{tr}\cos^2\theta^{}_{K^*} \nonumber\\
 & \hskip-0.6in + & \hskip-0.3in 
\sqrt{2}{\rm\,Re}(A^*_\parallel A^{}_0)\sin^2\theta^{}_{tr}\sin 2\phi^{}_{tr}\sin2\theta^{}_{K^*} \nonumber \\
 & \hskip-0.4in - & \hskip-0.2in 
\eta\sqrt{2}{\rm\,Im}(A^*_0 A^{}_\perp)\sin 2\theta^{}_{tr}\cos\phi^{}_{tr}\sin2\theta^{}_{K^*} \nonumber \\
 & \hskip-0.2in - & \hskip-0.1in 
\left. 2\eta {\rm\,Im}(A^*_\parallel A^{}_\perp)
	\sin 2\theta^{}_{tr}\sin\phi^{}_{tr}\sin^2\theta^{}_{K^*}\,\right]\,,
\label{eqn:transversity}
\end{eqnarray}
where \azero, \apar, and \aperp\ are the complex amplitudes of the
three helicity states, and $\eta=+1\,(-1)$ for $B^0\,(\overline{B^0})$ 
decays. The amplitude \azero\ denotes the longitudinal polarization of 
the final state, and \aperp\ (\apar) denotes the transverse
polarization along the $z\,(y)$ axis. Note that 
$\left|A^{}_0\right|^2 + | A^{}_\parallel |^2 + \left|A^{}_\perp\right|^2 = 1$.
The value of $\left|A^{}_\perp\right|^2$ 
($1-\left|A^{}_\perp\right|^2=\left|A^{}_0\right|^2 + | A^{}_\parallel |^2$) 
is the \cp-odd (\cp-even) fraction of the decay. 

The complex amplitudes \azero, \aperp, and \apar\ are determined via an unbinned
ML fit to the candidates within the $m^{}_{bc}$-$\Delta E$ signal region;
the probability density function for signal is given by 
Eq.~(\ref{eqn:transversity}). By convention,
the value of Arg(\azero) is set to zero and $| A^{}_\parallel |^2$
is calculated from the normalization constraint. The results of
the fit are:
$\left|A^{}_0\right|^2 =0.43\,\pm0.09\,\pm\,0.04$, 
$\left|A^{}_\perp\right|^2 =0.41\,\pm0.10\,\pm\,0.04$, 
Arg(\apar)\,$= -2.57\,\pm\,0.39\,\pm\,0.09$, and 
Arg(\aperp)\,$= 0.48\,\pm\,0.32\,\pm\,0.06$.
The projections of the fit are shown in Fig.~\ref{fig:phikst_pol}.
The systematic errors include the (slow) pion detection efficiency
(3--6\%) and background from higher $K^*$ states (6--9\%).
The value of $\left|A^{}_\perp\right|^2$ obtained indicates that
both \cp-odd and \cp-even components of \bphikst\ are sizable.

\section{$B^\pm\rightarrow\rho^\pm\rho^0$ decays}

The decay \brhorho\ proceeds via $b\rightarrow d$ loop and 
$b\rightarrow u$ tree diagrams and contains two vector mesons 
in the final state. Angular correlations among the decay 
products ($\pi^+\pi^0\pi^+\pi^-$) 
can be used to search for \cp- and $T$-violating effects. 
In the final state, both $\rho$'s are either longitudinally 
or transversely polarized; the corresponding amplitudes are 
denoted \hzerzer\ and \honeone, respectively.

In this analysis, $\rho^+\rho^0$ states are reconstructed by combining 
three charged pions with one neutral pion. The charged pions are 
required to have $p^{}_T>0.10$\gevp.
Candidate $\pi^0$'s are reconstructed from $\gamma\gamma$ pairs
having $118<m^{}_{\gamma\gamma}<150$\mevm; each $\gamma$ must 
also satisfy $E^{}_\gamma>50\,(100)$\meve\ in the barrel (endcap) 
region. Candidate $\rho$ mesons
are identified via $\pi^+\pi^-$ or $\pi^+\pi^0$ pairs having
$0.65<m^{}_{\pi\pi}<0.89$\gevm. \brhorho\ candidates are identified
by requiring $5.272<m^{}_{bc}<5.290$\gevm\ and $-0.10<\Delta E<0.06$\geve.
The \hzerzer\ amplitude gives rise to asymmetric \rhopipi\ decays, i.e.,
one pion has high momentum and the other has low momentum. The \honeone\ 
amplitude gives rise to symmetric \rhopipi\ decays. Thus, the \hzerzer\ 
state has a lower reconstruction efficiency and a $\Delta E$ 
resolution $\sim$\,15\% broader than that for \honeone.

There are large backgrounds due to \qqbar\ continuum events. To 
reduce these we cut on both \Rqq\ and the thrust angle $\theta^{}_{thr}$,
which is the angle between the thrust axis of tracks originating from 
the $B$ candidate and that of the remaining tracks in the event. The cut 
chosen is $\left|\cos\theta^{}_{thr}\right|<0.80$. The overall rejection 
of continuum events is $>$\,99.5\%, with a signal efficiency of~28\%.
There is also a small level of background from $b\rightarrow c$ 
processes and rarer $B$ decays such as 
$B^+\rightarrow \eta'\rho^+,\,K^{*\,+}\rho^0,\,\rho^+K^{*\,0}$ and $\rho\pi$;
these tend to be displaced in $\Delta E$.

\begin{figure}
\hbox{
  \includegraphics[height=.27\textheight]{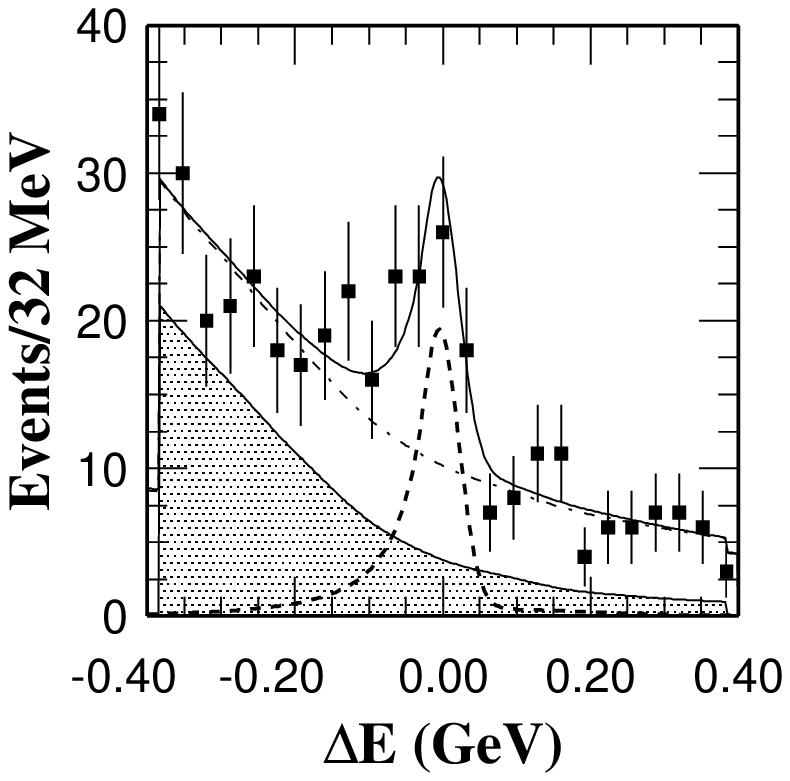}
\hskip0.30in
  \includegraphics[height=.27\textheight]{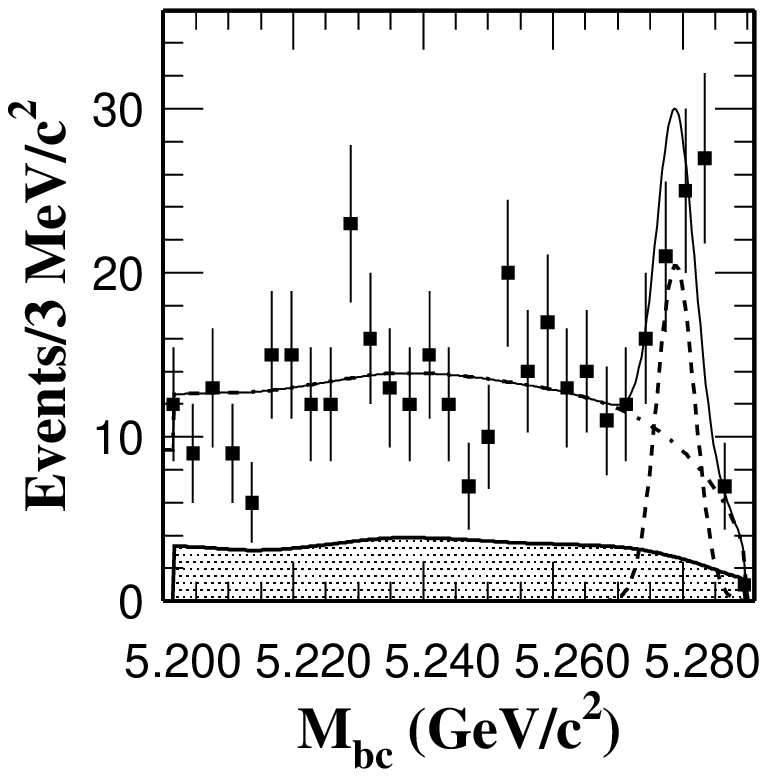}
}
\caption{$\Delta E$ (left) and $m^{}_{bc}$ (right) fits for candidate 
\brhorho\ decays. The shaded curve represents $B\bar{B}$ background,
the dash-dotted curve represents the sum of $B\bar{B}$ and continuum 
backgrounds, the dashed curve represents the \brhorho\ signal, and the 
solid curve represents the overall sum.
\label{fig:rhorho_signal}}
\end{figure}

\begin{figure}
\includegraphics[height=.26\textheight]{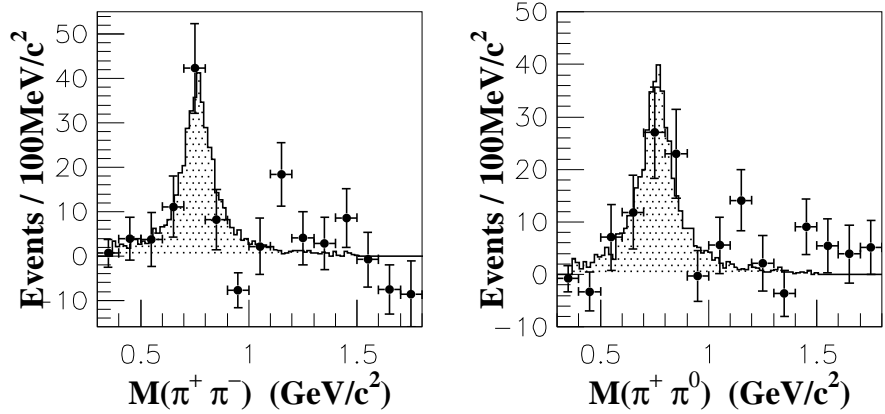}
\caption{$m^{}_{\pi^+\pi^-}$ (left) and $m^{}_{\pi^+\pi^0}$ (right)
distributions for candidate \brhorho\ decays. The shaded histogram
shows MC signal.
\label{fig:rhorho_rho}}
\end{figure}

The resulting $\Delta E$ and $m^{}_{bc}$ distributions are shown 
in Fig.~\ref{fig:rhorho_signal}. The event yields are determined 
by fitting in $\Delta E$. The fit yields $58.7\,\pm\,13.2$ events
with a statistical significance ($\sqrt{2\ln\left[L^{}_{max}/L^{}_0\right]}$\,)
of~5.3. Fitting the $\Delta E$ distributions for different $m^{}_{\pi\pi}$
bins gives the event yields plotted in Fig.~\ref{fig:rhorho_rho}. 
These distributions agree well with MC expectations and show little
nonresonant background beneath the $\rho$ peaks.

\begin{figure}
\hbox{
  \includegraphics[height=.21\textheight]{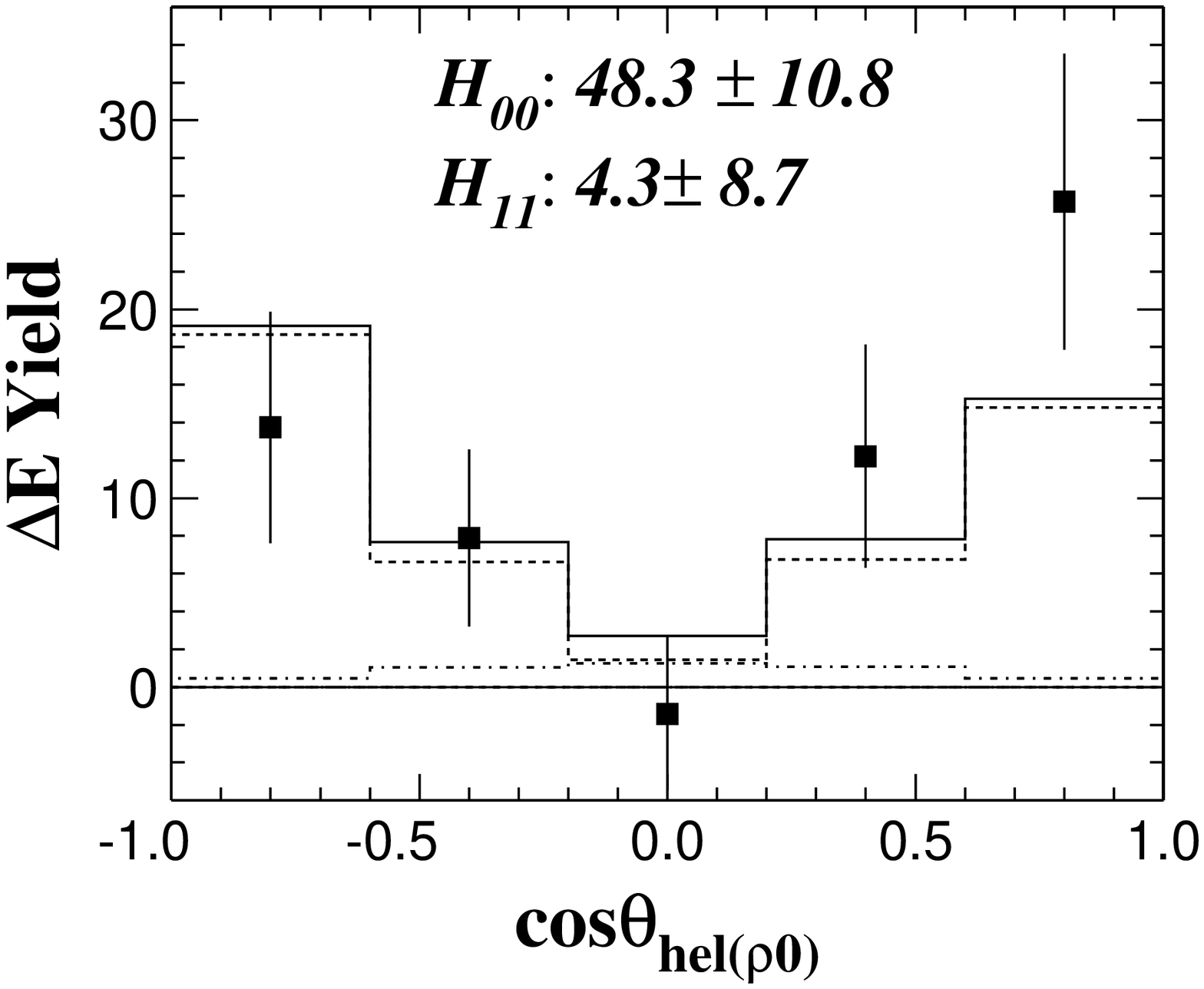}
\hskip0.30in
  \includegraphics[height=.21\textheight]{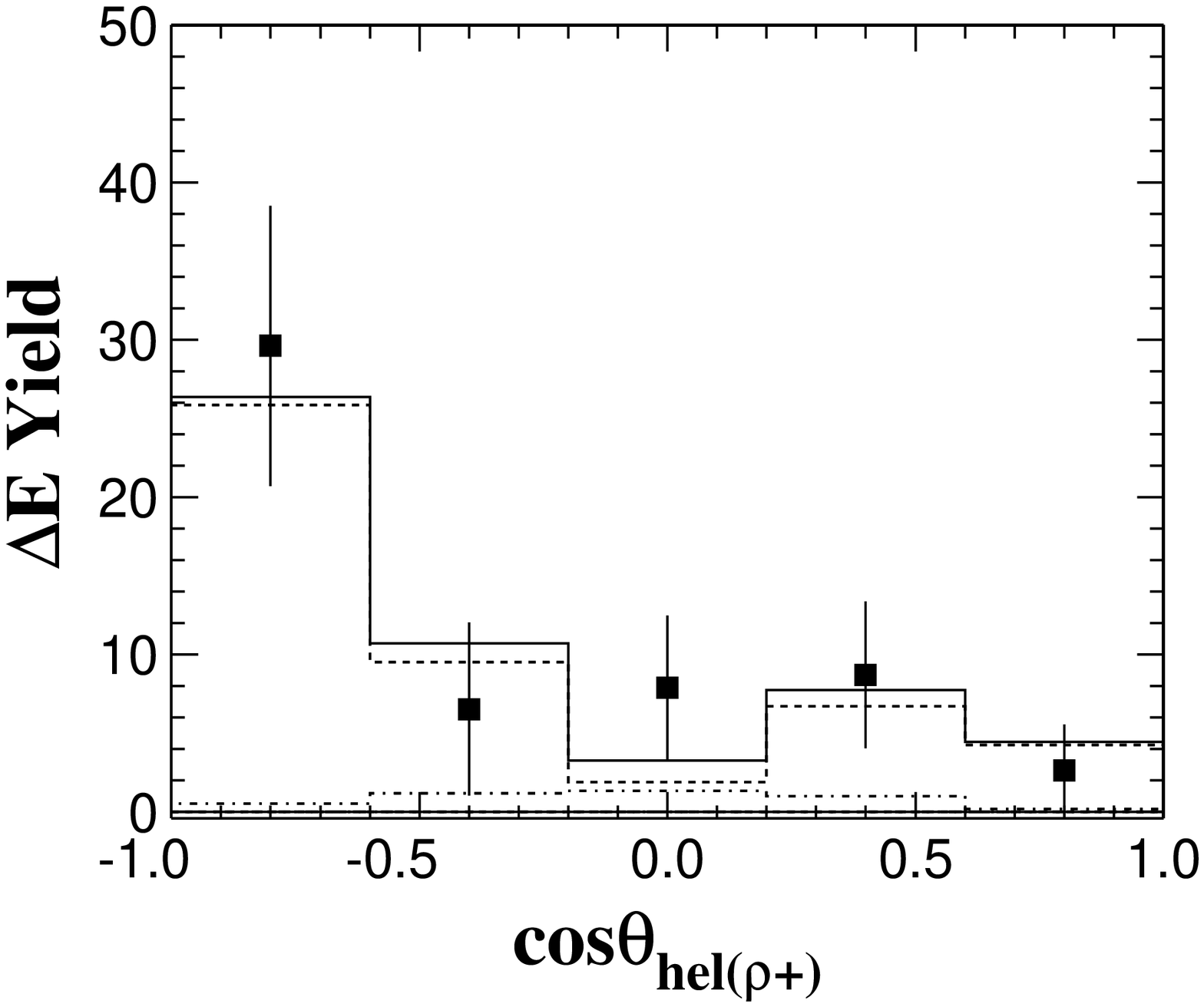}
}
\caption{Helicity distributions for the $\rho^0$ (left) and 
$\rho^+$ (right) in \brhorho\ decays. The dashed (dash-dotted)
histogram shows the \hzerzer\ (\honeone) component of the fit;
the solid histogram is their sum. The bin size is~0.4. The 
low event yield near $\cos\theta^{}_{hel(\rho^+)}=1$ is due 
to a requirement $p^{}_{\pi^0}({\rm CM})>0.5$\gevp.
\label{fig:rhorho_hel}}
\end{figure}

The relative strengths of \hzerzer\ and \honeone\ are determined by 
studying distributions of the helicity angle \thetahel, which  
is the angle between the $\rho$ flight direction in the $B$ rest frame
and the $\pi^+$ flight direction in the $\rho$ rest frame. The signal
yields determined from $\Delta E$ fits for different $\cos\theta^{}_{hel}$
bins are plotted in Fig.~\ref{fig:rhorho_hel} for both the $\rho^0$ and $\rho^+$.
We perform simultaneous binned fits to these distributions using MC
expectations for the \hzerzer\ and \honeone\ helicity states.
The fit yields the fraction of $\rho^0\rho^+$ final states that
are longitudinally polarized:
\begin{eqnarray*}
\frac{\Gamma^{}_L}{\Gamma^{}_{tot}} & = & (94.8\,\pm\,10.6\,\pm\,2.1)\,\%\,.
\end{eqnarray*}
This result shows that the \hzerzer\ state dominates, which is consistent
with theoretical expectations~\cite{long_dom}. The systematic error includes 
uncertainties in signal yield extraction and the polarization dependence of 
the detection efficiency. 
Based on this polarization ratio and the MC-determined reconstruction
efficiencies of the two helicity states, we calculate 
$B(B^+\!\rightarrow\!\rho^+\rho^0) = 
	\left(3.17\,\pm\,0.71\,^{+0.38}_{-0.67}\right)\times 10^{-5}$.

We separate the candidate events into $B^-\rightarrow\rho^-\rho^0$ 
and $B^+\rightarrow\rho^+\rho^0$ subsamples and fit these subsamples 
individually. The resulting event yields are $29.3\,\pm\,9.5$ and 
$29.3\,\pm\,9.1$, respectively. The \cp\ asymmetry is 
$A^{}_{CP}= [N(\rho^-\rho^0)-N(\rho^+\rho^0)]/[N(\rho^-\rho^0)+N(\rho^+\rho^0)]
=0.00\,\pm\,0.22\,\pm\,0.03$, which is consistent with zero.

\section{Summary}

With 78~fb$^{-1}$ of data the \belle\ experiment has:
\begin{itemize}
\item updated the branching fractions and \cp\ asymmetries for
$B^0\rightarrow\pi\pi$, $B^0\rightarrow K\pi$, and
$B^0\rightarrow KK$ decays;
\item measured the \cp\ asymmetries in \bdcpk\ decays, where \dcp\ 
represents a $D^0$ decaying to a $CP=+1$ or $CP=-1$ eigenstate, 
and investigated the possibility 
of using the measured asymmetries to constrain the CKM phase $\phi^{}_3$;
\item updated the branching fractions and \cp\ asymmetries for 
$B^\pm\rightarrow\omega\pi^\pm$ and $B^\pm\rightarrow\omega K^\pm$ decays;
\item measured the branching fractions for $B\rightarrow\phi K$ and
$B\rightarrow\phi K^*$ decays and, for the latter, measured the 
polarization amplitudes \aperp, \apar, and \azero\ in the transversity basis;
\item measured the branching fraction for \brhorho\ and the helicity amplitudes
\hzerzer\ and \honeone. This is the first reported observation of this decay.
\end{itemize}
Most results are consistent with theoretical expectations, although some
channels show interesting and possibly important discrepancies at the 
$2\sigma$ level. We look forward to investigating these further 
(and refining all of these measurements) with more data.

\bibliographystyle{aipproc}   

\bibliography{cipanp}

\end{document}